\documentclass{ws-ijmpe}
\usepackage[super,compress]{cite}
\usepackage{xcolor}

\begin{document}

\markboth{M. D. Cozma and W. Trautmann}{Neutron Star Radii from Laboratory Experiments}


\title{NEUTRON STAR RADII FROM LABORATORY EXPERIMENTS
}

\author{M. D. COZMA}

\address{IFIN-HH, Reactorului 30, 077125 M\v{a}gurele-Bucharest, Romania\\
dan.cozma@theory.nipne.ro}

\author{W. TRAUTMANN}

\address{GSI Helmholtzzentrum f\"{u}r Schwerionenforschung GmbH\\
Planckstr. 1, D-64291 Darmstadt, Germany\\
w.trautmann@gsi.de}

\maketitle

\begin{history}
\received{07 March 2025}
\end{history}

\begin{abstract}  
Our present knowledge of the nuclear equation of state is briefly reviewed in this article intended for a wider readership.
Particular emphasis is given to the asymmetric-matter equation of state required for modeling neutron stars, neutron-star mergers,
and r-process nucleosynthesis. Recent analyses based on combining information obtained from nuclear theory, heavy-ion 
collisions and astrophysical observations confine the obtained radii of the canonical 1.4-solar-mass 
neutron star to values between 12 km and 13 km. The remaining uncertainty is primarily related to missing information in
the density interval between nuclear saturation density and about twice that value which, however, 
is accessible with laboratory experiments. 
\end{abstract}



\section{Introduction}
\label{sec:intro}

The first observation of a merger of two neutron stars by the gravitational wave detectors of 
the LIGO (Laser Interferometer Gravitational-Wave Observatory) and Virgo Collaborations in August 2017, event GW170817, 
marked the beginning of a new era of multi-messenger astronomy including 
gravitational wave detection.\cite{abbott17,apj848,lattimer21} 
The analysis of the observed gravitational wave signal and of the associated electromagnetic emissions has provided us 
with the first clear demonstration that r-process nucleosynthesis occurs in neutron star mergers\cite{watson19} and, 
at the same time, with a new type of constraint for the nuclear equation of state. 
The tidal effects inscribed into the inspiral part of the gravitational waves emitted before contact
reflect the deformabilities of the stars and are related to their radii.\cite{abbott18,chaves19} 
The radii, in turn, are governed
by the balance between the gravitational forces compressing the stars and the internal pressure forces counteracting 
them.\cite{lattprak07,lipr08} 

The Neutron Star Interior Composition Explorer (NICER) was installed at the International Space Station in the same year 2017. 
From precise measurements of the pulse profiles of the two millisecond pulsars J0030+0451 and J0740+6620, 
two independent analysis groups based in Amsterdam and Maryland extracted values
for their radii with accuracies of approximately $\pm 1$~km.\cite{riley19,miller19,riley21,miller21} 
In particular, the simultaneous knowledge of the radius and mass of the heavy pulsar PSR J0740+6620 
(PSR for Pulsating Source of Radio) serves as a strong constraint for models constructed to describe neutron-star matter.
 
\begin{figure}[ht]		
\centerline{\includegraphics[width=0.70\columnwidth]{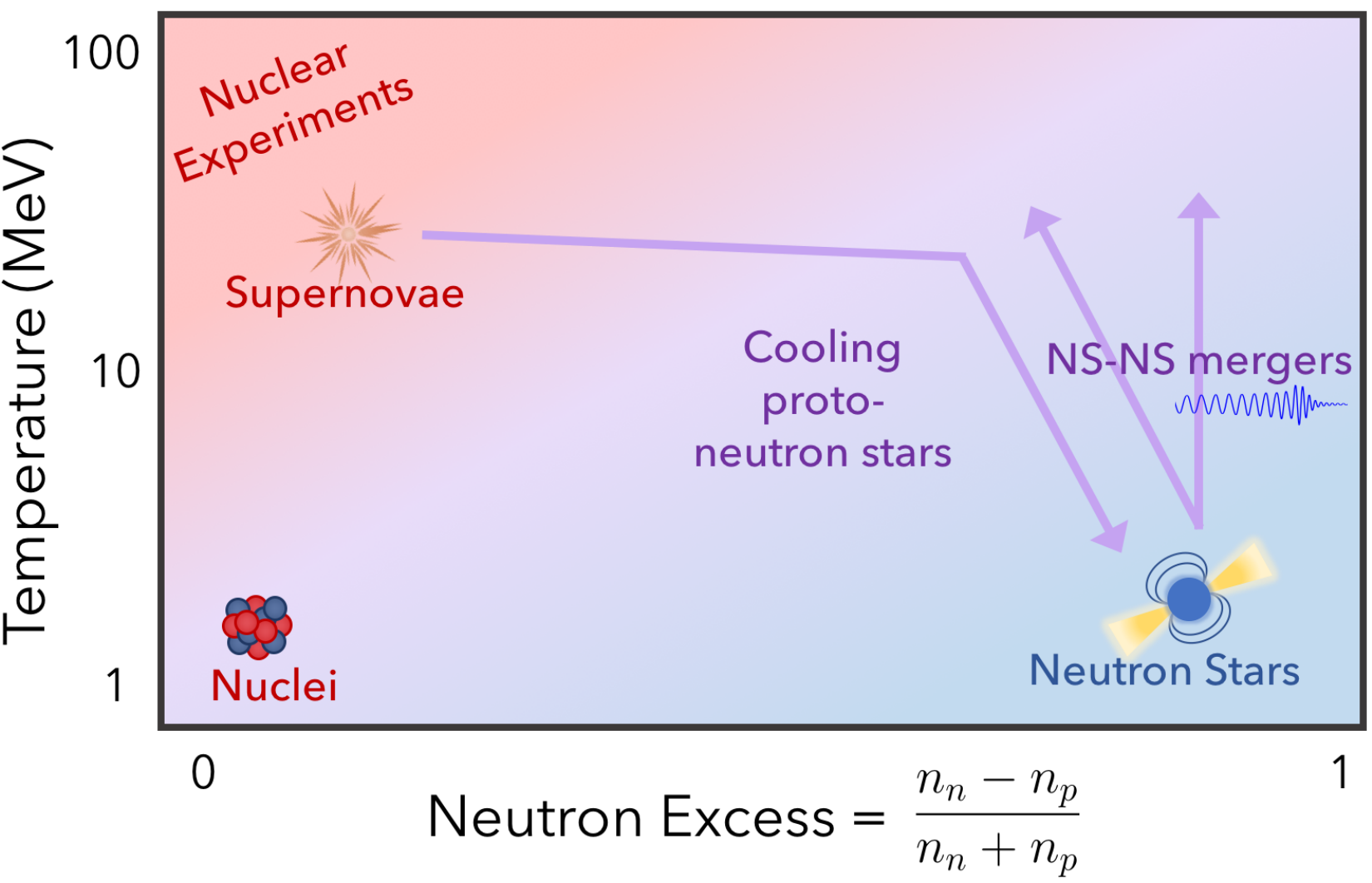}}
  \caption{Schematic representation of the temperature versus asymmetry regimes associated with the sources of 
   information available for the study of the equation of state of nuclear matter and of neutron-star matter. 
   The quantities $n_n$ and $n_p$ represent the neutron and proton densities, respectively. The arrows are meant to
   indicate the dynamical processes of neutron star formation and of neutron star (NS) collisions with other compact objects 
   (from Ref.\protect\cite{raithel19}, \textcopyright~2021 AAS reprinted with permission).
}
\label{fig:raithel}
\end{figure}

The sources of information now available for confining the nuclear equation of state as a function of temperature
and neutron excess (equal to the isotopic asymmetry $\delta = (n_n - n_p) / (n_n + n_p)$) are distributed over 
all four corner sections of the schematic Fig.~\ref{fig:raithel} taken from the 
work of Raithel, \"{O}zel, and Psaltis.\cite{raithel19}  
The figure emphasizes the different regimes of temperature and
asymmetry probed with terrestrial nuclear experiments and astrophysical observations. The abscissa, reaching 
from symmetric atomic nuclei to asymmetric neutron stars, in addition, stretches over the 55 orders of magnitude 
in mass that distinguish these objects under study.
It will be seen how amazingly well our techniques for describing nuclear forces can be applied to the different regimes indicated
in the figure. 

Nuclear structure deals with atomic nuclei in their ground and excited states and is thus limited to nuclear matter 
near and around the nuclear saturation density $\rho_0 = 2.7 \times 10^{14}$~g cm$^{-3}$ (corresponding to a nucleon
number density of $n_{\rm sat} = 0.16$~fm$^{-3}$; in the following, both $n$ and $\rho$ will be used as notation for 
densities given in numbers of nucleons per fm$^{3}$). Because of the contributions of surface effects, information 
provided by nuclear structure studies is most precise for densities around two thirds of that value. 
Nuclear reactions probe nuclear matter also at the higher and lower densities that parts of the collision system may pass through
for a short time. In heavy-ion collisions, densities of several times the saturation value may be reached in the 
central volume of the collision zone, depending on the collision energy.\cite{li_npa02} To exploit these opportunities, 
measurable signatures have to be found that characterize the modified matter at this particular instant. 
Neutron star properties, on the other hand, are the result of the balance between the gravitational forces and the 
counteracting pressure forces generated by the interacting constituents inside the star. The masses, radii or deformabilities 
of neutron stars are the result of the combined effects at all densities present inside the star. 
The so-called mass-radius relation is, therefore, a representation of the neutron-star-matter equation of state that is
equivalent to the usual form of the pressure as a function of density.\cite{lattprak07,lattimer12,baym18}

The arrows in Fig.~\ref{fig:raithel} are meant to indicate the trajectories in the temperature versus asymmetry plane
of the dynamical processes of neutron star formation\cite{prakash97,pons99} following the explosion of massive stars and of mergers of
neutron stars with other neutron stars or black holes, as now observed with gravitational waves.\cite{LIGOcatalog22}
To identify their role as sites for nucleosynthesis, a preeminent goal of nuclear and astro-physics,
knowledge of the equation of state is required over virtually the full
regime of temperature and asymmetry covered in the figure. A framework for treating the temperature dependence was proposed 
by Raithel {\it et al.}\cite{raithel19} 

This article will focus on the cold matter equation of state. 
Ultimately, as one hopes and believes, combining information from all sources with their relative weights will lead
to a unique solution determined with the statistically robust method of Bayesian inference.\cite{dietrich20} 
The individual ingredients will be introduced with brief 
characterizations of the methodical steps. All four domains indicated in Fig.~\ref{fig:raithel} contribute significantly and 
their mutual consistency is, in fact, quite impressive. Improvements regarding the accuracy of this 
type of results are necessary and possible. As it will become evident, a major role can be played by new information covering
the density interval between once and twice the saturation value which is accessible with laboratory experiments.

\section{Neutron Stars and Neutron Star Mergers}  
\label{sec:astro}

\begin{figure}[ht]		
\centerline{\includegraphics[width=0.75\columnwidth]{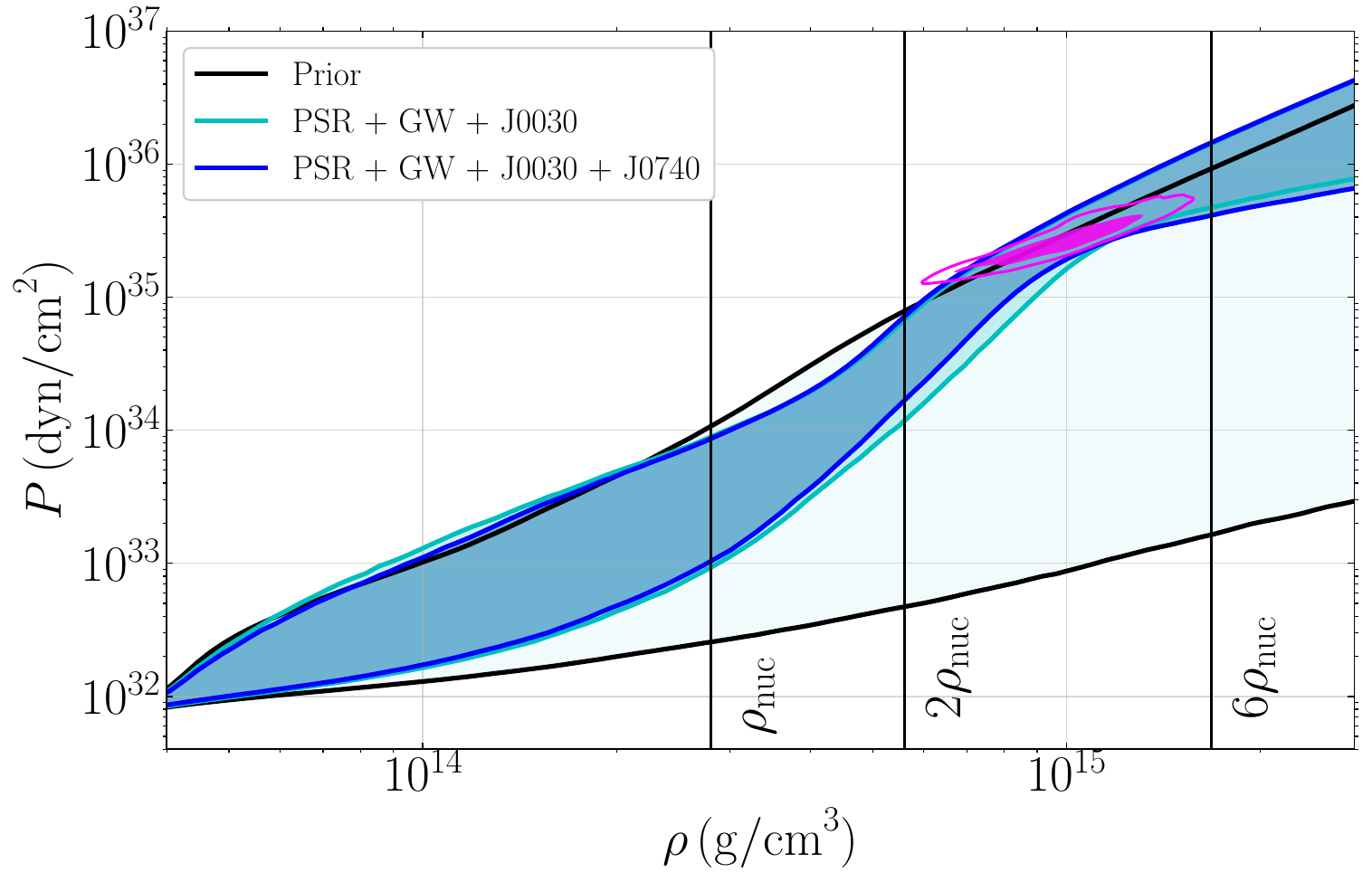}}
  \caption{Pressure-density relation of neutron-star matter: the shaded regions enclose the 90\% symmetric credible intervals. 
   Magenta contours give the 50\% and 90\% level of the
   central pressure-density posterior for the pulsar J0740+6620 inferred from all astrophysical data used for the analysis
   (reprinted with permission from Ref.\protect\cite{legred21}, copyright \textcopyright~2021 by the American Physical Society).
}
\label{fig:legred}
\end{figure}

An example of the equation of state of neutron-star matter as it can be obtained from astrophysical observations alone
is shown in Fig.~\ref{fig:legred}. The figure
is taken from the work of Legred {\it et al.}\cite{legred21} which was completed after the results of the 
radius measurement for J0740+6620, the presently heaviest pulsar (abbreviated from pulsating radio source) with a precisely 
known mass, have become available.\cite{cromartie19,fonseca21} 
It represents the obtained equation of state in the form of the pressure versus density relation and illustrates the information 
provided by the masses (labeled PSR in Fig.~\ref{fig:legred}), the radii, and the tidal deformability (GW) of the observed neutron stars. 
The magenta contours located at densities 
around 10$^{15}$~g/cm$^3$, roughly four times the saturation value, 
indicate the 50\% and 90\% credible intervals of the central pressure-density posterior 
for J0740+6620 inferred from all data used in the analysis. 
Note that 10$^{35}$ dyn/cm$^2$ corresponds to 10$^{29}$ bar, evidently a high pressure. 

The importance of the radius measurement for the heavy pulsar 
J0740+6620 is highlighted by indicating, in addition, the contours obtained without it (shown in green). 
The contours including it (in blue) provide a lower bound for the pressure at densities near and above twice
the saturation value that is significantly higher than that obtained by only using the radius measured for the
$\approx$1.4 solar-mass pulsar J0030+0451.\cite{riley19,miller19} At lower densities, the astrophysical observations do not 
contain information exceeding that of the model-agnostic process used for constructing the adopted prior 
distribution of trial solutions for the equation of state.\cite{legred21}

The most precise masses of neutron stars in our galaxy have been obtained for binary systems consisting of two neutron stars and are
narrowly distributed around a value of 1.4 solar masses.\cite{lattprak07} 
It has, therefore, become customary to express the stiffness of a
neutron-star-matter equation of state with its result for a 1.4 solar-mass neutron star. Larger radii indicate stiffer
solutions. Legred {\it et al.} report a
radius $R_{1.4} = 12.6 \pm 1.1$~km with the error representing the 90\% confidence limit.\cite{legred21} The value 
$R_{1.4} = 11.9 \pm 1.4$~km resulting from the revised analysis of the LIGO/Virgo Collaborations\cite{abbott18}, 
mainly based on the observed tidal deformability, is smaller due to the lower pressures  
indicated in Fig.~\ref{fig:legred} but consistent within errors.
Evidently, stiffer solutions were favored by the observation of neutron stars with masses close 
to and above two solar masses. Their existence is known since 2010 when Demorest {\it et al.} 
reported a value of $1.97 \pm 0.04$
solar masses for J1614-2230 which represented the first case of a precisely measured
mass of this magnitude.\cite{demorest10} With the method of measuring Shapiro delay, used also for
J0740+6620 with $2.08 \pm 0.07$ solar masses,\cite{cromartie19,fonseca21} the gravitational
effect of the compact companion star by delaying the radio signals of the pulsar is determined. 
The mass of the companion star deduced therefrom then permits resolving the ambiguity of the mass of the neutron star
in the orbital equations. 
For the third pulsar J0348+0432 in this class with a mass of $2.01 \pm 0.04$ solar masses, the method of phase-resolved
optical spectroscopy applied to its white-dwarf companion was used.\cite{antoniadis13}

A radius $R_{1.4} = 12.45 \pm 0.65$~km (68\% credibility), similar to that of Legred {\it et al.}\cite{legred21} 
but with a slightly smaller uncertainty, was 
obtained by Miller {\it et al.} by including information related to lower densities.\cite{miller21}
The authors argue that up to about half the saturation density, roughly corresponding to the core-crust transition density
in neutron stars, the equation of state is well known.\cite{douchin01,tews17,baym19} The contribution to the radius uncertainty 
related to modeling the crust should not exceed 300~m
(we note here that the structure of the crust itself, containing extremely neutron-rich nuclei\cite{wolf13} and unusual forms of 
nuclear matter, and its possible excitations in the form of vibrations are very interesting topics 
in their own right\cite{baym18,chamel08}).
For densities above half the saturation value, several carefully constructed models were used 
in this analysis which, however, have not been
validated by laboratory measurements. How strongly the conclusions depend on the class of equation-of-state models 
that is employed and how these choices influence the results and their errors is taken into account.\cite{miller21}

The highest densities that may be reached in the core of heavy neutron stars (Fig.~\ref{fig:legred}) has attracted 
particular attention because of the possibility of a phase transition to quark matter in the central core of the star. 
The density at which this might occur and whether it is likely to be confirmed by observation is presently a matter of debate. 
It is being argued that the maximum density may still be too low\cite{brandes23} but it is as well demonstrated that 
current observational data are compatible with quarkyonic matter existing in the core of massive stars.\cite{tang21}
(for further reading on this subject see, e.g., Refs.\cite{weih20,soma23,annala23,pang24}).

\section{Precise Information from Nuclear Structure}
\label{sec:precise}

\begin{figure}[ht]		
\centerline{\includegraphics[width=0.65\columnwidth]{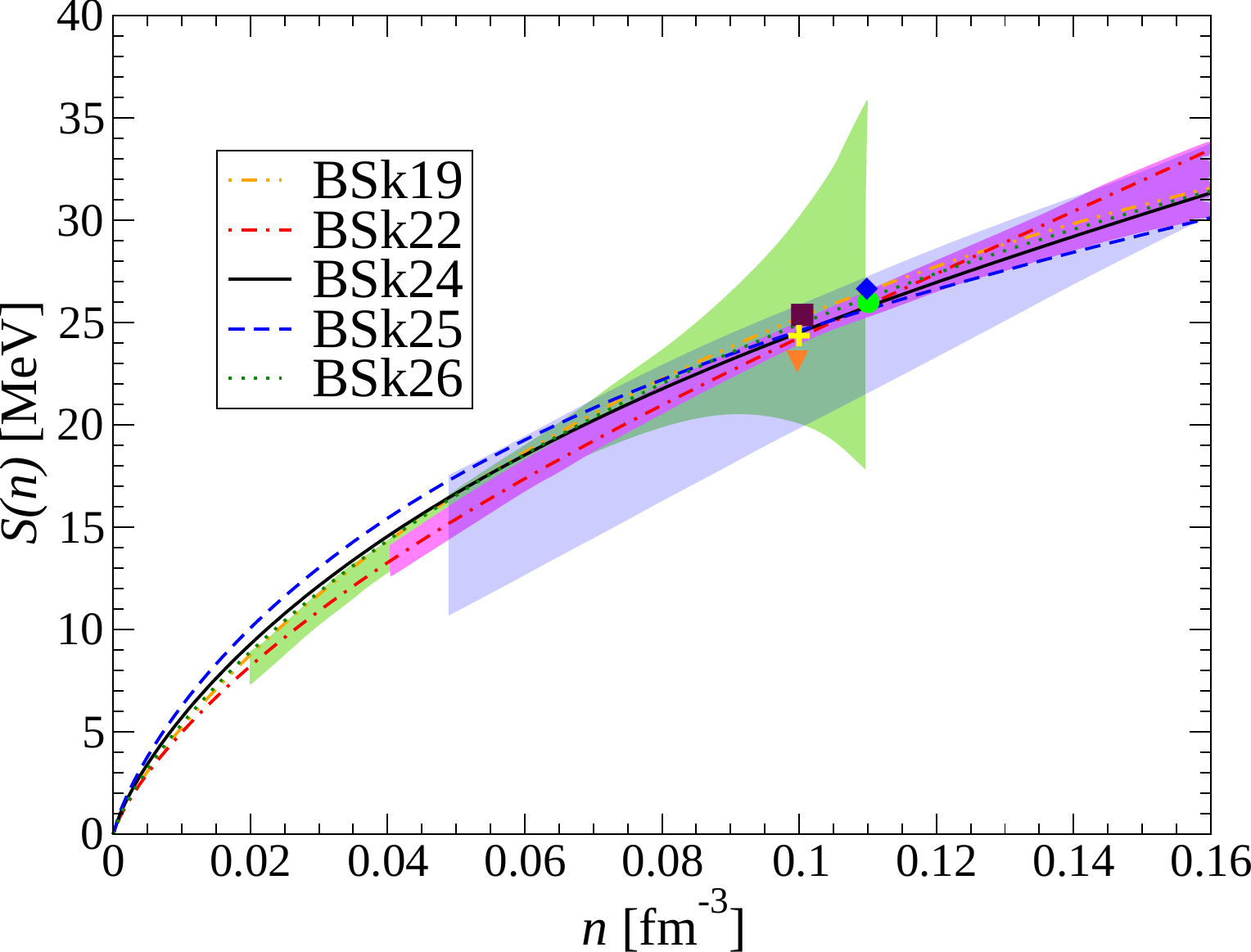}}
  \caption{Variation of the nuclear-matter symmetry energy $S(n)$ as a function of the nucleon density $n$ 
for the Brussels-Montreal functionals. The shaded areas are
experimental constraints from the electric dipole polarizability of $^{208}$Pb (green, Ref.\protect\cite{zhang15}),
heavy-ion collisions\protect\cite{tsang09} (blue), and from isobaric-analog states and neutron 
skins\protect\cite{dani14} (purple).
The symbols represent the results given in line 1 to 5 of Table~\ref{tab_1} in the sequence cross,\protect\cite{trippa08} 
triangle,\cite{rocamaza13} circle,\cite{wangou13} diamond,\cite{zhang13}
and square,\cite{brown13} respectively 
(reprinted with permission from Ref.\protect\cite{chamel19}, copyright \textcopyright~2019 by the American Physical Society).
}
\label{fig:chamel}
\end{figure}

The ground-state properties of atomic nuclei and the structure of their excitations document the nuclear equation of state 
at zero or very low temperatures. Because of the finite volumes of the surface regions, the average densities probed 
are below the saturation value encountered in the center of heavy nuclei.\cite{khan13} 
Perot, Chamel and Sourie, in their report on the Brussels-Montreal equations of state,\cite{chamel19}
cite five references with explicitly given values for the density to which the value obtained 
for the symmetry energy applies.\cite{trippa08,rocamaza13,wangou13,zhang13,brown13}. These results are listed 
in Table~\ref{tab_1} and displayed in Fig.~\ref{fig:chamel} reprinted from their paper. 

\begin{table}[ht]
\begin{center}
\begin{tabular}{|c|c|c|c|}
\hline
$\rho$ (fm$^{-3}$) & $E_{\rm sym}(\rho)$ (MeV) & Observable & Ref. \\ 
\hline
0.10 & $24.1 \pm 0.8$ & giant dipole resonance & \cite{trippa08} \\
0.10 & $23.3 \pm 0.6$ & giant quadrupole resonance & \cite{rocamaza13} \\
0.11 & $26.2 \pm 0.7$ & Fermi-energy difference & \cite{wangou13} \\
0.11 & $26.7 \pm 0.1$ & binding energy difference& \cite{zhang13} \\
0.10 & $25.5 \pm 1.0$ & binding energy and radius & \cite{brown13} \\
\hline
\end{tabular}
\end{center}
\caption{Results for the symmetry energy $E_{\rm sym}(\rho)$ with 1-$\sigma$ uncertainties at the specified nucleon 
densities $\rho$ as obtained from the analysis of nuclear structure data using selected force models. 
}
\label{tab_1}
\end{table}

The symmetry energy for nuclear matter, $E_{\rm sym} (\rho)$, can be defined as the coefficient of the quadratic 
term in an expansion of the energy per particle in the asymmetry 
$\delta = (\rho_n-\rho_p)/\rho$, where $\rho_n, \rho_p,$ and $\rho$
represent the neutron, proton, and total densities, respectively,
\begin{equation}
E/A(\rho,\delta) = E/A(\rho,\delta = 0) + E_{\rm sym}(\rho) \delta^2 + \mathcal{O}(\delta^4).
\label{eq:e_sym}
\end{equation}

\noindent The symmetry energy is a function of the density $\rho$ and, in the quadratic approximation,\cite{wen21} 
equal to the difference between the energies of symmetric matter ($\delta = 0$) and neutron matter ($\delta = 1$).
The densities $\rho$ listed in the first column of Table~\ref{tab_1} are close to 2/3 of the saturation density and the values for 
$E_{\rm sym}(\rho)$ scatter around 25~MeV. 
A most probable value for density $\rho = 0.10$~fm$^{-3}$ may be calculated by scaling the values listed in lines 3 and 4 
of the table to this density with the slope parameter $L$(0.11 fm$^{-3}) = 47 \pm 8$~MeV (Ref.\cite{zhang14}), 
leading to $24.8 \pm 0.7$~MeV and $25.2 \pm 0.3$~MeV, respectively.
A $\chi^2$ analysis yields $E_{\rm sym}(0.10$~fm$^{-3}) = 24.8 \pm 0.5$~MeV, documenting that the symmetry energy 
at this density is consistently determined by nuclear structure, as obtained with these 
studies based on different observables and analysis methods. 

The fact that nuclear structure offers precise information for a narrow density interval has been known for many 
years (see, e.g., Refs.\cite{brown00,fuchs06}) and has served as motivation for studies aiming at other density regimes. 
Zhang and Chen report that their result for the electric dipole polarizability of $^{208}$Pb provides precise information 
for densities up to about one half of the saturation value.\cite{zhang15} It is shown in Fig.~\ref{fig:chamel} together with 
results for the symmetry energy $S(n)$ obtained from the observation of isospin diffusion 
in heavy-ion reactions\cite{tsang09} and 
from analyzing isotopic analog states\cite{dani14} that, in this form, have been presented in the review of 
Horowitz {\it et al.}\cite{horowitz13} in 2013 (here, the alternative notation $S(n)$ is used 
for the symmetry energy $E_{\rm sym}(\rho)$). All three colored bands cover the value 24.8 MeV at 0.10~fm$^{-3}$ 
rather comfortably and, together, constitute a consistent result for the symmetry energy at subsaturation density. 

The phenomenon of isospin diffusion\cite{tsang09} during the fragmentation process following energetic heavy-ion collisions 
gives access to the isospin equilibration governed by the strength of the symmetry potentials as the fragments form and detouch 
from the excited collision system. These processes proceed through non-equilibrium stages at high temperatures and require 
analyses with transport models to deduce properties of the acting forces. Isobaric analog states are characterized by mutual 
exchanges of neutrons and protons in specific orbits. Their energy differences thus, in principle, consist of the 
differences of the symmetry term and of changes in the Coulomb term which have to be evaluated. 
Danielewicz and Lee concluded that the deduced constraints for the symmetry energy are valid for the density 
interval $0.04 \le \rho \le 0.13$~fm$^{-3}$ (Ref.\cite{dani14}). 
 
The energy-density functionals of the Brussels-Montreal series BSk19-BSk26, developed for the use in studies 
of neutron star physics, are also displayed in Fig.~\ref{fig:chamel}. 
Their parameters were primarily determined by fitting to the 2353 measured masses of atomic nuclei with $Z>8$ from the 
2012 Atomic Mass Evaluation\cite{audi12} and adjusted to reproduce individual neutron-matter equations of state.\cite{chamel19}
The five members of the series differ slightly in their predictions for $E_{\rm sym}(\rho_0)$, there assuming values between 
29 and 32 MeV, but differ considerably in their predictions at higher densities.\cite{chamel13} 
At density $\rho = 0.10$~fm$^{-3}$, they agree with each other 
within half a MeV and very precisely with the common value obtained above (Fig.~\ref{fig:chamel}). 

The tendency of mean-field models to approach $S(n) = 0$~MeV at very low density is no longer realistic once the clustering 
degree of freedom of symmetric nuclear matter is considered.\cite{horo06,typel10,nato10} Heavy-ion reactions in the Fermi 
energy regime proved to be useful for determining clustering probabilities and provided constraints 
for equation-of-state models for low-density matter with particular applications in 
supernova physics.\cite{hempel15,bougault20} 
A detailed account of how to calculate the equation of state in the different regions of a neutron star, 
and in particular for the low-density crust, is given by Perot, Chamel and Sourie in their report\cite{chamel19}
(for reviews see, e.g., Refs.\cite{baoan19,burgio21}).

\section{The Symmetry Energy at Saturation}
\label{sec:saturation}

It has been noticed by Li and Han several years ago that the many results obtained for the nuclear symmetry energy from 
terrestrial nuclear experiments and 
astrophysical observations are amazingly compatible, even though individual results may be afflicted
with considerable uncertainties.\cite{lihan13} The  compilation of Oertel {\it et al.} based on 53 
laboratory results and astrophysical observations has arrived at weighted averages $E_{\rm sym}(\rho_0) = 31.7 \pm 3.2$~MeV 
for the symmetry energy at saturation density $\rho_0$ and $L = 58.7 \pm 28.1$~MeV for the slope parameter $L$ describing 
its density dependence.\cite{oertel17}. The parameter $L$ is the coefficient of the linear term of the Taylor expansion of 
the symmetry energy with respect to the density relative to $\rho_0$

\begin{equation}
E_{\rm sym}(\rho)=E_{\rm sym}(\rho_0)+\frac{L}{3}\left(\frac{\rho-\rho_0}{\rho_0}\right)+
\frac{K_{\rm sym}}{18}\left(\frac{\rho-\rho_0}{\rho_0}\right)^2+ ...
\label{eq:taylor}
\end{equation}

\noindent and $K_{\rm sym}$ is the still fairly unknown coefficient of the quadratic or curvature 
term.\cite{lipr08,cozma18,baoan20} 

New information has been provided by terrestrial experiments aiming at determining the nuclear equation of state at 
densities closer to the saturation density encountered in the center of heavy nuclei. The thickness $\Delta R_{np}$ 
of the neutron skin
of $^{208}$Pb nuclei is considered a particularly valuable observable because it reflects the pressure experienced
by neutrons in the interior of this neutron-rich nucleus which directly relates to the slope parameter $L$ of the 
symmetry energy.\cite{horowitz01}  
Brown has demonstrated that the knowledge of the neutron skin of heavy nuclei permits the extrapolation 
from $\rho = 0.10$~fm$^{-3}$ 
to saturation density.\cite{brown13} Present experimental values of $\Delta R_{np} \approx 0.16$~fm for $^{208}$Pb, 
obtained from measuring the electric dipole reponse of heavy nuclei and reported to be accurate within 
$\approx \pm 0.03$~fm,\cite{rossi13,tamii14} correspond to uncertainties of more than $\pm 2$~MeV for 
the extrapolated $E_{\rm sym}(\rho_0) \approx 32$~MeV according to Brown.\cite{brown13} 
With the unexpectedly large central value of the thickness of the neutron skin of $^{208}$Pb, $\Delta R_{np} = 0.28 \pm 0.07$~fm,  
reported by the PREX-2 Collaboration,\cite{adhikari21,reed21} the same extrapolation will lead to values 
in the neighborhood of 40 MeV for $E_{\rm sym}(\rho_0)$. 

The PREX results were awaited with high expectation because the measured 
asymmetry of parity-violating electron scattering on $^{208}$Pb is interpreted without the aid of nuclear theory.\cite{horowitz20} 
Reed {\it et al.} deduce  $E_{\rm sym}(\rho_0) = 38.1 \pm 4.7$~MeV from this measurement.\cite{reed21} 
A result closer to currently favored values, $E_{\rm sym}(\rho_0) = 33.0 \pm 2.0$~MeV, 
has been obtained by Essick {\it et al.}\cite{essick21} by combining astrophysical data with PREX-2 and chiral effective 
field theory constraints (here and in the following, if reported error margins 
are not very different in positive and negative directions, only the larger value is quoted.)

A new situation
arose when the small value $\Delta R_{np} = 0.121 \pm 0.026 {\rm (exp)} \pm 0.024 {\rm (model)}$ for $^{48}$Ca was reported by 
the CREX Collaboration using the same method.\cite{adhikari22} Theory very consistently maintains that the two values should be
correlated, with each of the models predicting approximately equal skin thicknesses 
for $^{48}$Ca and $^{208}$Pb.\cite{adhikari22,reinhard22}
This puzzle is presently not resolved and a wide range of opinions have been expressed.
Within the quoted errors which are mainly statistical, 
the two results were shown to be not necessarily contradicting each other or 
theoretical expectations\cite{lattimer23,zhang23,zhou24} 
whereas, on the other hand, modified energy density functionals are being proposed to more closely accomodate the constraints 
imposed by the PREX and CREX measurements.\cite{reed24} New data with the high resolution predicted for the proposed
MREX experiment\cite{becker18} at Mainz will be very useful here.

Because of the valuable information carried by a precise value for the thickness of the neutron skin of $^{208}$Pb, 
alternative methods are of particular interest. Hu {\it et al.} reported on {\it ab-initio} calculations leading to 
$\Delta R_{np} = 0.17 \pm 0.03$~fm,\cite{hu22} whereas Giacalone {\it et al.} deduced a value 
$\Delta R_{np} = 0.22 \pm 0.06$~fm from the analysis of the $^{208}$Pb+$^{208}$Pb collisions at ultrarelativistic 
energy performed at the Large Hadron Collider.\cite{giacalone23} 
Both measurements are compatible with the earlier nuclear structure data\cite{rossi13,tamii14}
but do not necessarily reduce the overall uncertainty.
Smaller errors within $\pm 0.02$~fm were predicted for measurements of neutron-removal cross sections in high-energy nuclear 
collisions of 0.4 to 1 GeV/nucleon, provided that the applied reaction theories can be sufficiently tested and 
constrained in a necessary series of separate measurements.\cite{aumann17,ponnath24}
For further reading on the nuclear symmetry energy see, e.g., the review of Baldo and Burgio\cite{baldo16} and
the {\it Topical Issue on Nuclear Symmetry Energy}
published by The European Physical Journal A in 2014.\cite{epja2014}

\section{Heavy-Ion Collisions}
\label{sec:HIC}

Heavy-ion collisions (HIC) at sufficiently high energy provide the possibility of exploring the density interval between once 
and twice the saturation value $\rho_0$, even though only for short times of the order of 30 fm/$c$ ($10^{-22}$~s) 
or less.\cite{li_npa02,li02,dani02} Information regarding the nuclear equation of state is extracted from
observed strengths of collective flows and from cross sections for meson production. 
Collective particle motion refers to anisotropies of particle yields 
caused by pressure gradients within the compressed and heated matter 
whereas meson yields may reflect density dependent collision rates. To the extent that these observations can be
related to early dynamical stages of the collision, they will provide access to pressure and high density, the quantities 
of interest. Theoretical reaction models are required for quantitative interpretations.

\begin{figure}[ht]		
\centerline{\includegraphics[width=0.65\columnwidth]{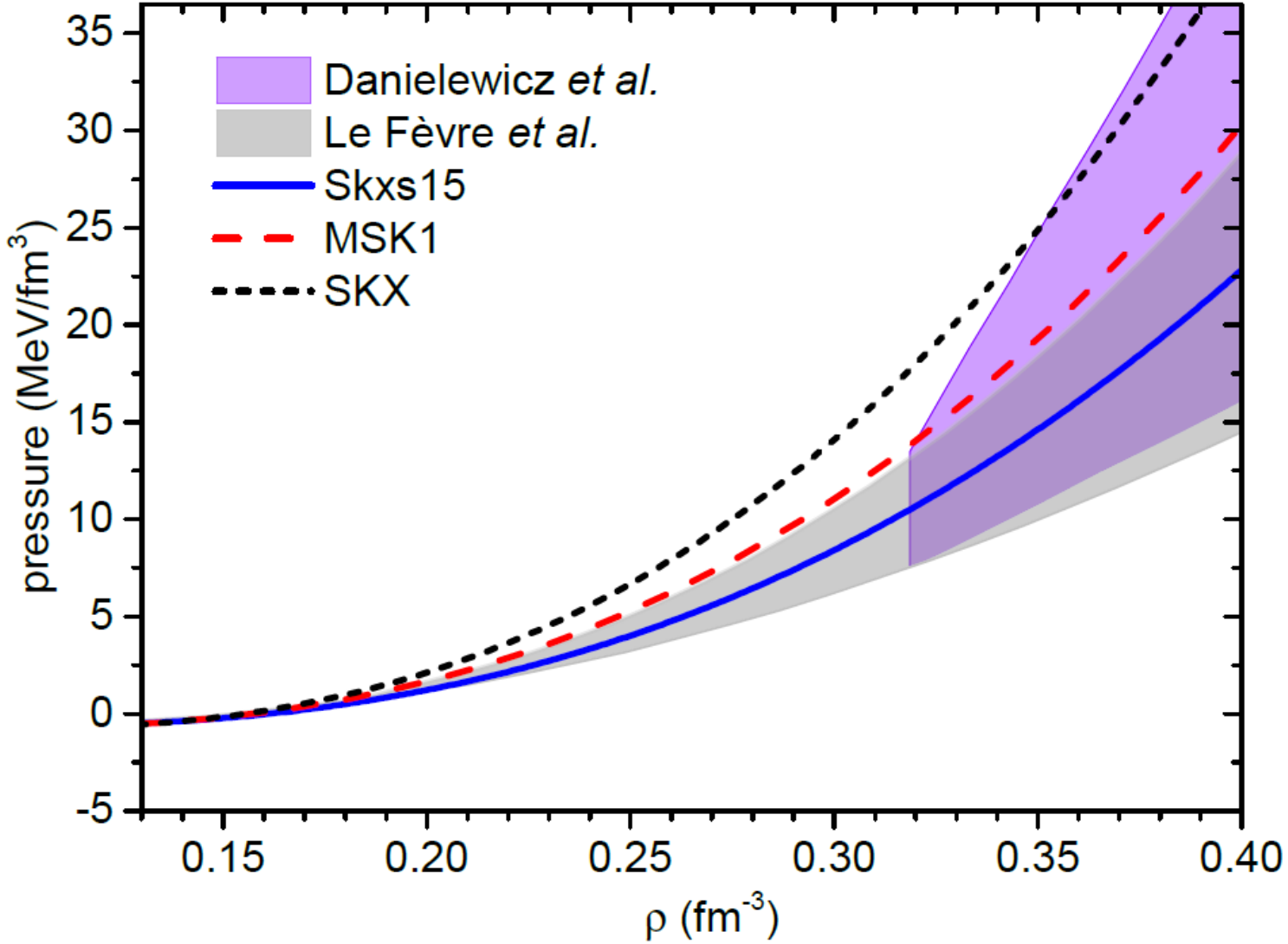}}
  \caption{Pressure in symmetric nuclear matter as a function of density. The lines represent predictions for the 
   Skxs15 (solid blue line), MSK1 (dashed red line), and SKX (dotted black line) interactions. The analysis of 
   Wang {\it et al.}\protect\cite{wang18} favors the interval between Skxs15 and MSK1. 
   The shaded regions represent the results 
   obtained by Danielewicz {\it et al.}\protect\cite{dani02} and Le F\`{e}vre {\it et al.}\protect\cite{lefevre16} 
   (from Ref.\protect\cite{wang18}).
}
\label{fig:wang}
\end{figure}

Heavy-ion collisions have been important for symmetric matter by establishing the soft nature of the equation of state 
(nuclear incompressibility $K_0 \approx 200$~MeV) and the momentum dependence of the acting nuclear 
forces.\cite{fuchs01,dani00} 
The collective flows of nucleons in $^{197}$Au+$^{197}$Au collisions observed at the Bevalac accelerator at
Lawrence Berkeley National Laboratory and the Alternating Gradient
Synchrotron (AGS) at Brookhaven National Laboratory at energies up to 11.5 GeV/nucleon were analyzed 
by Danielewicz {\it et al.}\cite{dani02} and the results converted into a 
pressure band for symmetric nuclear matter at densities reaching up to $4.5 \rho_0$. The covered large density interval
made the data useful for comparisons with astrophysical results.\cite{steiner13}
A comprehensive data set of the collective observables directed and elliptic flow has been presented by the Four-Pi (FOPI) 
Collaboration for a variety of collision systems in the SIS energy range up to 1.5 GeV/nucleon.\cite{reisdorf10,reisdorf12}
The maximum densities are between 2
and $3 \rho_0$. The recent analyses of the elliptic flow results measured for $^{197}$Au+$^{197}$Au collisions performed by 
Le F\`{e}vre {\it et al.}\cite{lefevre16} and Wang {\it et al.}\cite{wang18} match very well the AGS-deduced pressures 
at $2 \rho_0 = 0.32$~fm$^{-3}$ (Fig.~\ref{fig:wang}).

The forces used are conventionally characterized by the incompressibility $K_0$ describing the variation of the pressure 
for small amplitudes in density but contain higher-order terms. 
In the case of Le F\`{e}vre {\it et al.},\cite{lefevre16} the reported incompressibility is $K_0 = 190 \pm 30$ MeV. 
If the third-order term is assumed to be responsible for the pressures of 7.5--13 MeVfm$^{-3}$ obtained for $2 \rho_0$, 
its value is of the order of $Q_0 \approx -260 \pm 60$~MeV.

Information on the asymmetric matter equation of state is expected from differential observables that are specifically sensitive to
the pressure and density differences experienced by protons and neutrons in the high-density zone of the collisions.
Different pressures acting on neutrons and
protons in asymmetric matter at high density may thus be recognized as isotopic variations of collective 
flows\cite{li_baoan00,russotto11,cozma11} and yields.\cite{kaneko21}
Meson production at threshold in two-step processes via $\Delta$ production is favored at higher densities 
because of the 
higher collision rates among the constituents in the high-density zone.
To the extent that the isotopic composition there is influenced by the asymmetric matter equation of state,
differential effects can be expected for observables as, e.g., kaon or pion ratios.\cite{ferini06} 
The interpretation of measured data relies on transport theory accounting for the non-equilibrium dynamics of the reaction.

\begin{figure}[ht]		
\centerline{\includegraphics[width=0.90\columnwidth]{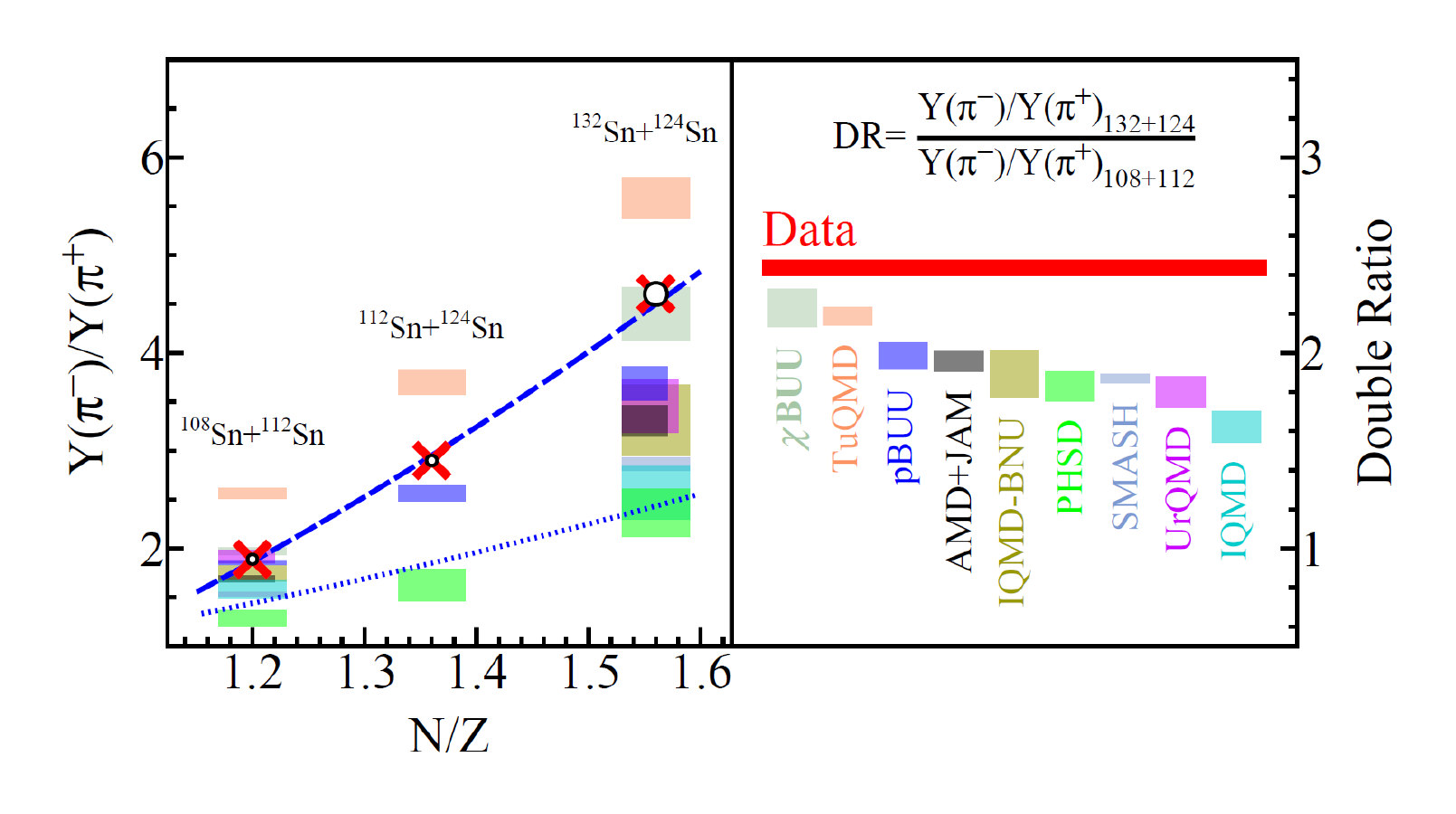}} 
\vskip -0.1cm  
  \caption{Left panel: charged-pion yield ratios from the S$\pi$RIT experiments\protect\cite{jhang21} as a function 
   of the combined $N/Z$ of the studied Sn+Sn collision systems at 270 MeV/nucleon.  
   The data are displayed as red crosses with the size of the open 
   symbols inside representing the experimental errors. The results of the calculations are shown as colored boxes for 
   the different codes identified by their color in the right panel. The height of the boxes is given by the difference of 
   predictions for the soft and stiff choices of the symmetry energy. The dashed blue line is a power-law fit with the 
   function $(N/Z)^{3.6}$, whereas the dotted blue line represents $(N/Z)^2$ of the system. 
   Right panel: double ratios (DR) of pion yields for $^{132}$Sn+$^{124}$Sn 
   and $^{108}$Sn+$^{112}$Sn. The data and their uncertainty are given by the red horizontal bar and the results of the 
   transport models are shown as colored boxes, in a similar way as in the left panel
   (please see Refs.\protect\cite{jhang21,wolter22} for descriptions of the codes and references; 
   reprinted from Ref.\protect\cite{wolter22}, with permission from Elsevier).
}
\label{fig:jhang}
\end{figure}

Pion production in heavy-ion collisions has been extensively investigated by the FOPI Collaboration and a comprehensive set of 
production yields and yield ratios of charged pions has been collected.\cite{reisdorf07}
More recent studies indicate, however,
that the interpretation of the integrated pion ratios cannot be considered as 
conclusive at the present time (see, e.g., Refs.\cite{hong14,songko15,cozma16,lili2017} and references 
given therein).   
The S$\pi$RIT Collaboration has more recently continued the investigation of pion production and employed
radioactive $^{108}$Sn and $^{132}$Sn beams for extending the isotopic compositions of the studied Sn+Sn reaction systems on 
the neutron poor and neutron rich side.\cite{jhang21} Central collisions at an incident energy 270 MeV/nucleon close
to the pion production threshold were selected.

\begin{figure}[ht]		
\centerline{\includegraphics[width=0.95\columnwidth]{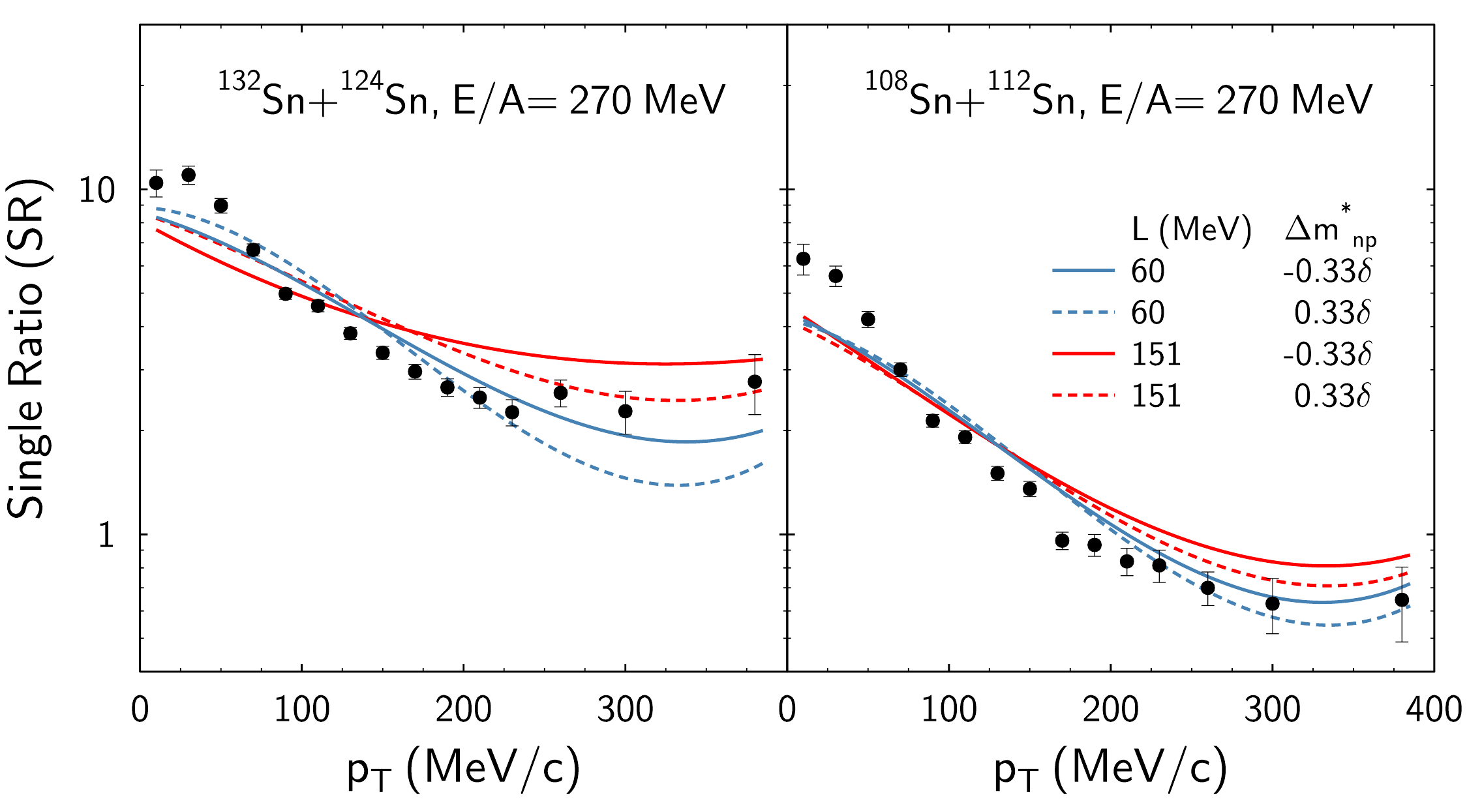}}
  \caption{Single pion spectral ratios for $^{132}$Sn+$^{124}$Sn  (left) and $^{108}$Sn+$^{112}$Sn (right) reactions. 
   The curves are dcQMD predictions using different $L$ and $\Delta m^*_{np}$ values listed in the right panel
   (adapted from Ref.\protect\cite{estee21}, copyright \textcopyright~2021 by the American Physical Society).
}
\label{fig:estee}
\end{figure}

Also here the conclusion was reached that the differences of model predictions for the momentum-integrated yield ratios and 
double ratios are larger than their sensitivities to the density dependence of the symmetry energy as illustrated in
Fig.~\ref{fig:jhang}, an extension of the figure reported in the original paper.\cite{jhang21} The right panel, in particular,
presents the results of nine transport model predictions, finalized before the experimental results were known,
for the double ratio (DR) of charged pions from the neutron rich and neutron poor reaction systems.
The height of the colored symbols represents the difference of the predictions for the soft and stiff choices of the
symmetry energy. Deviations from the experimental results are in nearly all cases larger than the
calculated sensitivity to the stiffness of the symmetry energy. Even in the case of the $\chi$BUU\cite{zhangko18} code which
comes closest to the experimental result, the latter is not covered by the sensitivity interval of the model. 
Investigations of possible sources of differences among transport models\cite{xu24} and between several transport models and 
experiment have appeared subsequently.\cite{cozma21,ikeno23}

The investigation of spectral instead of integrated yield ratios has shown, however, 
that the high-energy part of the spectra responds reliably to changes in the assumptions for the isovector part 
of the forces used in the transport description, even though it remains dependent on the chosen value for the 
effective mass splitting in the asymmetric medium (Fig.~\ref{fig:estee}).\cite{estee21} 
The obtained result 42 MeV $< L < 117$~MeV is compatible with previous knowledge of the the density dependence of the 
symmetry energy but carries a large uncertainty, to a large part also related to the fading experimental 
accuracy at high transverse momentum.
The description of pion production near threshold touches at the limit of what is presently possible with the semi-classical 
transport approach for describing the temporary evolution of the collision including resonances. The
Transport Model Evaluation Project (TMEP) is a community-wide effort to enhance the reliability of model predictions.
A review containing short descriptions of the codes of the participating groups has appeared recently.\cite{wolter22}

\section{Elliptic Flow Ratios}
\label{sec:elliptic}

Results obtained by studying collective elliptic flows in heavy-ion collisions were included in the first study of the
neutron-star-matter equation of state that combined information from 
astrophysical observations of neutron stars, neutron-star mergers, and from collisions of gold nuclei at
relativistic energies with microscopic nuclear theory calculations.\cite{huth22}
In particular, information from the FOPI\cite{reisdorf12,lefevre16} and the Asymmetric-Matter EOS (ASY-EOS) experimental
campaigns\cite{russotto16} were used to obtain new constraints for neutron-rich matter at densities around 
saturation up to $2 \rho_0$. 

The FOPI results, together with the Bevalac and AGS data, were discussed in the previous section (Fig.~\ref{fig:wang}). 
The ASY-EOS experiments had been motivated by the analysis of the earlier FOPI-LAND data\cite{leif93,lamb94} which 
demonstrated the usefulness of elliptic-flow ratios and elliptic-flow differences of neutrons versus protons and 
neutrons versus charged particles for gaining information on the density dependence of the nuclear 
symmetry energy.\cite{russotto11,cozma11,traut12, wang14}
Systematic effects influencing the collective flows of neutrons and charged particles in similar ways 
should cancel in the ratios or differences and thus enhance the relatively small asymmetry effects in the moderately 
asymmetric $^{197}$Au+$^{197}$Au collision system. It was also shown that the model dependence of the obtained 
results is small.\cite{cozma13,russotto14}

\begin{figure}[ht]           
  \centerline{\includegraphics[width=0.55\columnwidth]{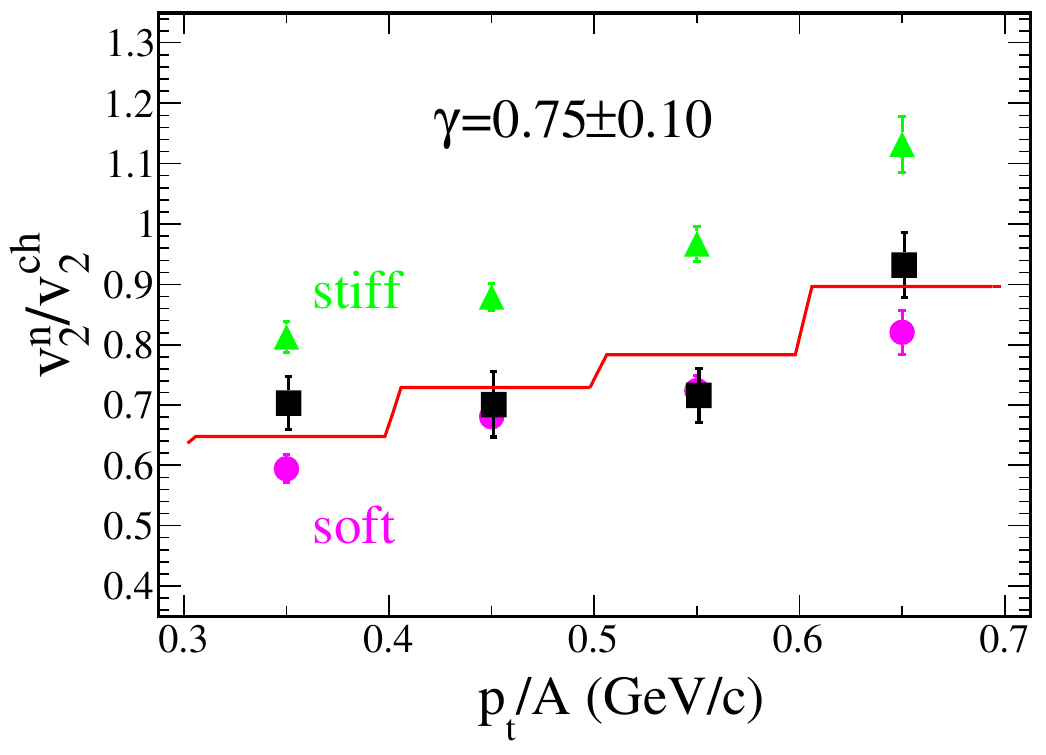}}
   \caption{Elliptic flow ratio of neutrons over all charged particles for central collisions of
$^{197}$Au+$^{197}$Au at 400 MeV/nucleon as a function of the transverse momentum per nucleon $p_t/A$. 
The black squares represent the experimental data, the green triangles and purple circles represent the
UrQMD predictions for stiff ($\gamma = 1.5$) and soft ($\gamma = 0.5$)
power-law exponents of the potential term, respectively. The
solid line is the result of a linear interpolation between the
predictions, weighted according to the experimental errors of
the included four bins in $p_t/A$, and leading to the indicated
$\gamma = 0.75 \pm 0.10$.
(reprinted with permission from Ref.\protect\cite{russotto16}, copyright \textcopyright~2016 by the American Physical Society).
}
\label{fig:russotto_fig14}       
\end{figure}

Motivated by these findings, an attempt was made to improve the 
accuracy with a new experiment conducted at the GSI laboratory in 2011 (ASY-EOS experiment S394). 
The experimental setup\cite{russotto16} followed the scheme developed for FOPI-LAND by 
using the Large Area Neutron Detector (LAND\cite{LAND}) placed at $\theta_{\rm lab} \approx 45^{\circ}$ 
as the main instrument for neutron and charged particle detection.
The strength of elliptic collective flows is quantitatively described with the second Fourier coefficient $v_2$
of the azimuthal distributions near mid-rapidity with respect to the event-by-event determined reaction plane
after corrections for the finite dispersion of the latter are applied.\cite{russotto16,andronic06}

As in the FOPI-LAND experiment, the reaction $^{197}$Au+$^{197}$Au was studied at 400 MeV/nucleon, close to the energy at
which collective emissions perpendicular to the reaction plane, also called squeeze-out, reach a maximum. 
Correspondingly, the $v_2$ parameter is negative and passes through a minimum.\cite{andronic06,lefevre18} 
Constraints for the symmetry energy were determined by comparing the ratios 
of the elliptic flows of neutrons and charged particles (ch), $v_2^{n}/v_2^{ch}$, with the corresponding model 
predictions for soft and stiff assumptions (Fig.~\ref{fig:russotto_fig14}).
For the analysis, a version of the UrQMD transport model adapted to the study of intermediate energy heavy-ion 
collisions was employed.\cite{qli11}
Different options for the dependence on isospin asymmetry were implemented,
expressed as a power-law dependence of the potential part of the symmetry energy on the
nuclear density $\rho$ according to

\begin{equation}
E_{\rm sym} = E_{\rm sym}^{\rm pot} + E_{\rm sym}^{\rm kin} 
= 22~{\rm MeV} (\rho /\rho_0)^{\gamma} + 12~{\rm MeV} (\rho /\rho_0)^{2/3}.
\label{eq:pot_term}
\end{equation}

\noindent Here $\gamma =0.5$ and $\gamma =1.5$ correspond to a soft and a stiff density 
dependence, respectively, as shown in Fig.~\ref{fig:russotto_fig14}. For the symmetry energy at saturation,
the parametrization assumes $E_{\rm sym}(\rho_0) = 34$~MeV.

\begin{figure}[ht]           
  \centerline{\includegraphics[width=0.58\columnwidth]{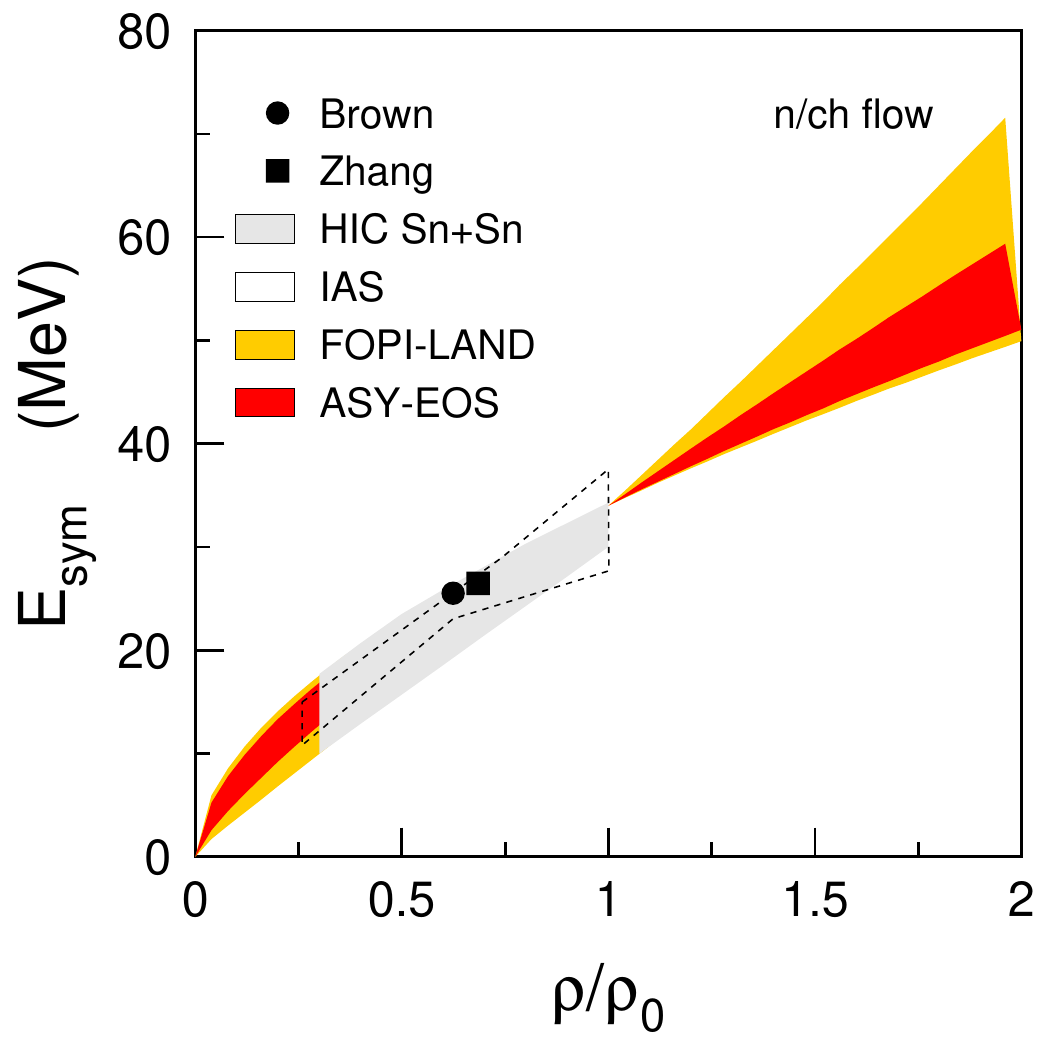}}
   \caption{Constraints deduced for the density dependence of the symmetry energy from the ASY-EOS 
data~\protect\cite{russotto16} in comparison with the FOPI-LAND result of Ref.~\protect\cite{russotto11} as a function 
of the reduced density $\rho/\rho_0$. 
The low-density results of Refs.~\protect\cite{brown13,zhang13,tsang09,dani14} as reported in Ref.~\protect\cite{horowitz13} 
and included in Fig.~\protect\ref{fig:chamel}
are given by the symbols, the grey area (HIC), and the dashed contour (IAS). For clarity, the FOPI-LAND and ASY-EOS 
results are not displayed in the interval $0.3 < \rho/\rho_0 < 1.0$
(reprinted with permission from Ref.\protect\cite{russotto16}, copyright \textcopyright~2016 by the American Physical Society).
}
\label{fig:russotto_fig18}       
\end{figure}

Applying the necessary corrections and their uncertainties converts the fit result of the acceptance-integrated 
elliptic-flow ratio (Fig.~\ref{fig:russotto_fig14}) into a power-law coefficient $\gamma = 0.72 \pm 0.19$. 
This is the result displayed as a red band in Fig.~\ref{fig:russotto_fig18} as a function of the 
reduced density $\rho/\rho_0$. It has a considerably smaller uncertainty than the earlier
FOPI-LAND result shown in yellow. The chosen parametrization appears compatible with the low-density 
behavior of the symmetry 
energy from Refs.\cite{brown13,zhang13,tsang09,dani14}, discussed in Section~\ref{sec:precise} and included in the figure.  
The slope parameter describing the variation of the symmetry energy with density at saturation
is $L = 72 \pm 13$~MeV. The corresponding curvature term in the Taylor expansion with respect to density is in the 
range $K_{\rm sym} = -70$ to -40~MeV.

The range of densities probed with the neutron versus charged particle elliptic-flow 
ratio was explored with the T\"{u}bingen version of the QMD model (T\"{u}QMD, Ref.\cite{cozma13}) by studying the effect
of systematically modifying the asymmetry term at different density intervals in a series of test calculations. It was found
that, at the given energy of 400 MeV/nucleon, the sensitivity curve is centered at saturation density but forms a broad distribution
from low densities up to about 2.5 times the saturation value.\cite{russotto16} The same T\"{u}QMD transport model was also
used for a comprehensive study of the parameter dependence of the calculated predictions.\cite{cozma13}

\section{Interpretation of Flow Data}
\label{sec:interp}

The nature and quality of the measured elliptic flow ratios (Fig.~\ref{fig:russotto_fig14}) does not permit fits with more than 
one free parameter which, according to Eq.~(\ref{eq:pot_term}), is the power law parameter $\gamma$. The value 
$E_{\rm sym}(\rho_0) = 34$~MeV and the power-law functional form of the potential term were assumptions and kept fixed. 
A choice $E_{\rm sym}(\rho_0) = 31$~MeV, close to the lower end of the presently favored interval
of $E_{\rm sym}(\rho_0)$, lowers the slope parameter to $L = 61 \pm 9$~MeV but the functional form of a power-law dependence 
on density of the potential term remains as an assumption.\cite{russotto16} 

Further analyses of the flow data appeared, some of them addressing the shortcomings just mentioned.
The work of Cozma\cite{cozma18} started from the so-called MDI force (MDI stands for momentum dependent interaction) 
originally designed to produce a fixed $E_{\rm sym}(\rho_0) = 31.6$~MeV and containing a parameter $x$ which controls
the density dependence of the symmetry energy.\cite{lipr08,das03} 
By adding a density dependent term, a new force MDI2 with an additional parameter $y$ was generated which, 
together with the original $x$, permitted independent choices for the slope and curvature parameters $L$ and $K_{\rm sym}$ of
Eq.~(\ref{eq:taylor}). By lifting the correlation of these two parameters, 
present in most models, uncertainties in the short-range three-body 
force are admitted and thereby taken into account.\cite{cozma18} The second modification concerned the density
at which the symmetry energy assumes a predetermined value. It was moved from the saturation density to the 
lower value 0.10~fm$^{-3}$ at which uncertainties are small (cf. Section~\ref{sec:precise}), and the value 
$E_{\rm sym}$ (0.10 fm$^{-3}$) = 25.5 MeV obtained by Brown\cite{brown13} was adopted.

\begin{figure}[ht]           
  \centerline{\includegraphics[width=0.65\columnwidth]{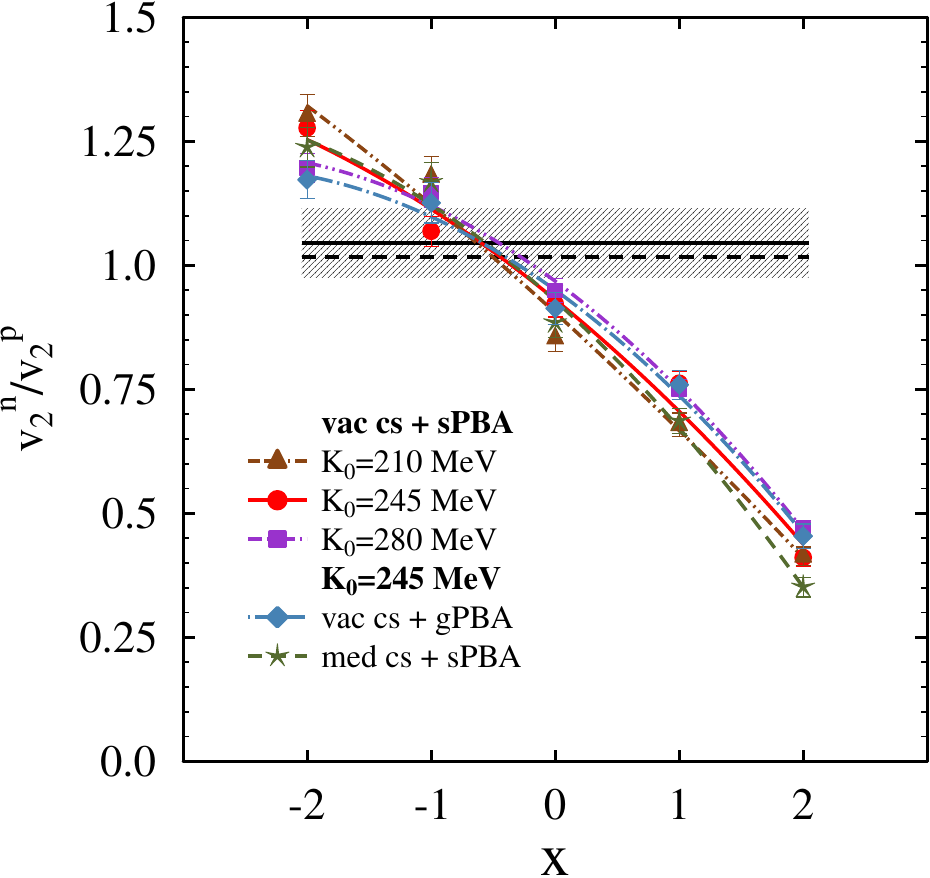}}
   \caption{Neutron-to-proton elliptic flow ratio as a function of the stiffness parameter $x$ in
comparison with the FOPI-LAND data. Experimental results as reported in Refs.~\protect\cite{wang14,cozma13}  
are indicated by the dashed and full horizontal lines, respectively, with only the error band corresponding 
to the latter data set being displayed (hashed). See text for explanations of the symbols and connecting lines
(from Ref.~\protect\cite{cozma18}, reprinted with permission from Springer Science+Business Media).
}
\label{fig:cozma3}
\end{figure}

The T\"{u}QMD, equipped with the MDI2 force and supplemented by a phase-space coalescence model adjusted to qualitatively 
describe FOPI experimental multiplicities of free nucleons and light clusters,\cite{reisdorf10} was used to compare the 
theoretical predictions with the FOPI-LAND neutron-to-proton and ASY-EOS neutron-to-charged-particles elliptic flow ratios.
Additional corrections that were applied are motivated and described in the paper.\cite{cozma18} 
The final result presented there is 
$L = 85 \pm 22({\rm exp}) \pm 20 ({\rm th}) \pm 12 ({\rm sys})$~MeV and 
$K_{\rm sym} = 96 \pm 315({\rm exp}) \pm 170 ({\rm th}) \pm 166 ({\rm sys})$~MeV.\cite{cozma18} The systematic errors
listed separately take, e.g., into account that the experimental light-cluster-to-proton multiplicity ratios were 
underestimated by the calculations. It was the achievement made in this work to obtain a solution up to the quadratic term 
in Eq.~(\ref{eq:taylor}) without assumptions, however at the cost of a larger uncertainty.

The largest contribution of the error is experimental and, as it turns out, mainly related to the neutron-proton flow 
ratio presently only available from the FOPI-LAND data\cite{russotto11}  
and thus affected by the limited statistics collected in that experiment. 
This can be improved with new measurements, so that much more precise values for the slope and curvature terms can 
be expected in the future. The example depicted in Fig.~\ref{fig:cozma3} illustrates this point, there for calculations with 
the original MDI force. It shows that various choices made for the symmetric-matter incompressibility $K_0$ (colored symbols
triangle, circle and square), the cross section parametrizations vac cs (vacuum cross sections) or med cs (in-medium
cross sections), or the algorithms simulating Pauli blocking (sPBA or gPBA), which all affect the calculated individual flows,
have only small impact on flow ratios. Smaller experimental errors would considerably improve the constraint on $x$ 
and thus on the density dependence of the symmetry energy. 

Very recent work with an improved version of the T\"{u}bingen QMD model (dcQMD\cite{cozma24})
has particularly addressed open questions regarding the in-medium 
elastic nucleon-nucleon cross-sections, accounting for their dependence on density and isospin asymmetry,
and the momentum dependence of the interaction adjusted by means of fixing nucleon effective masses at saturation density.
The study uses only stopping and flow data of the FOPI 
Collaboration\cite{reisdorf10,reisdorf12,andronic03,reisdorf04} without making use of
the differential observables neutron versus charged particles flow ratio or charged-pion ratio.
It builds on previous efforts demonstrating the existence of 
a hierarchy of in-medium effects on nucleonic observables,\cite{dani02} the largest being due to in-medium modifications 
of cross-sections, followed by those due to the momentum dependence of the interaction and the density dependence
of the equation of state.

\begin{figure}[ht]		
\centerline{\includegraphics[width=0.70\columnwidth]{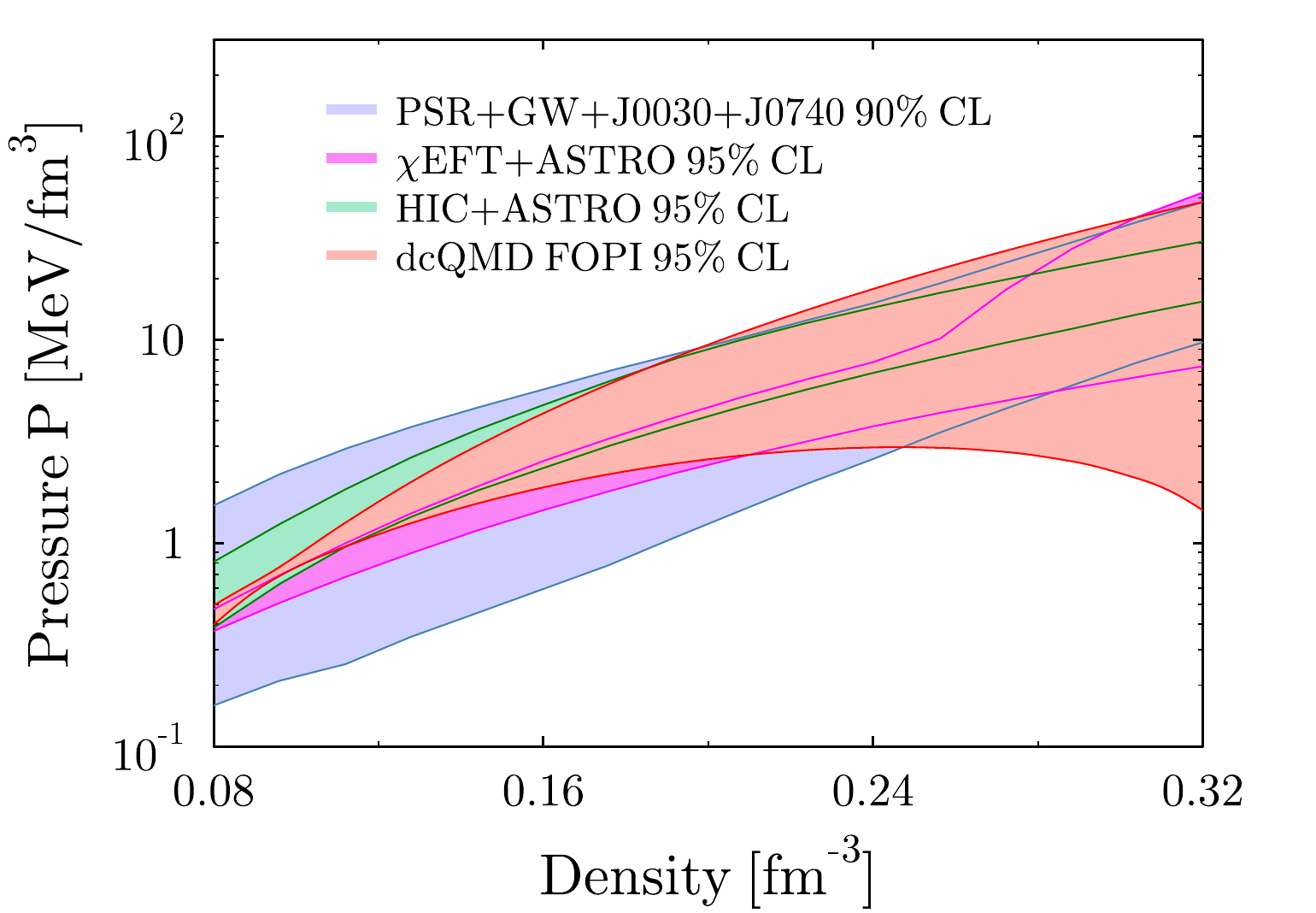}}
  \caption{Pressure of cold neutron-star matter (asymmetry $\delta = 0.93$) as a function of the baryon 
number density as obtained in Ref.\protect\cite{cozma24} with the dcQMD model (red) in comparison with the results of 
Legred {\it et al.} (blue),\protect\cite{legred21} Huth {\it et al.} (magenta),\protect\cite{huth22}
and Tsang {\it et al.} (green)\protect\cite{tsang24} 
(adapted from Ref.\protect\cite{cozma24}, copyright \textcopyright~2024 by the American Physical Society).
}
\label{fig:cozma_fopi}
\end{figure}

It was recognized in this study that the dependence of in-medium cross-sections on isospin asymmetry has a non-negligible 
impact and, for that reason, comparisons with experimental data for systems with a wide range of neutron-to-proton ratios were crucial.
Incident energies were restricted to $E_{\rm lab} \le 800$~MeV/nucleon to reduce uncertainties related to resonance production
and propagation. As in the previous analysis,\cite{cozma18} the symmetry energy at two thirds saturation density
is kept fixed as $E_{\rm sym}$(0.10 fm$^{-3}$) = 25.5 MeV  but, to reduce the number of free parameters, $K_{\rm sym}$ and 
higher order terms were taken as correlated with $L$.
The obtained pressure versus density contour of cold neutron-star matter with 95\% credibility is
given in Fig.~\ref{fig:cozma_fopi} (in red) and compared to three combined analyses to be introduced in the section below.
It represents a rather narrow constraint for densities up to around the saturation value.
Besides the result $K_0 = 230 \pm 11$~MeV for 
the compression modulus of symmetric matter, a constraint $L = 63 \pm 13$~MeV for the slope parameter 
describing the density dependence of the symmetry energy is obtained. 
As an important result, the study documents that valuable and
consistent information is contained in the isotopic dependences of charged particle cross sections and flow data.
Their sensitivity to density is found to be strongest near and below saturation but to also extend beyond $2 \rho_0$.

\section{Combined Analyses}
\label{sec:combined}

Results that combine the information from astrophysical observations with nuclear physics results including heavy-ion collisions 
have been reported recently. Bayesian inference has become the standard framework for considering the available sources
of information with their proper statistical weights. In each step of a sequential procedure, new information is used to 
inform the so far obtained distribution of candidate equations of state, the prior of that step, with the new input. 
The resulting posterior distribution then serves as the new prior in the following step.

The procedure starts with an initial prior distribution which has to be carefully chosen because it affects the final outcome. 
So-called agnostic priors are obtained by choosing large intervals with wide open borders for the parameter set of
the chosen equation-of-state model. The term metamodeling\cite{margueron18} applies to models based on the Taylor expansion 
given in Eq.~(\ref{eq:taylor}). A distribution of candidate equations of state is generated by systematically varying
the expansion coefficients. It remains a challenge to keep the computing demands within realizable limits because, 
for each of the candidates, the probability of their reproducing the set of observations will have to be calculated.
 
Chiral effective field theory ($\chi$EFT) is an alternative choice for generating a prior distribution, 
based on the expectation that it represents our knowledge of the nuclear forces at low energies.\cite{drischler21a} 
$\chi$EFT is a systematic expansion in powers of a typical momentum scale $p$ over the $\chi$EFT breakdown scale
$\Lambda_b$. For infinite matter, $p$ is of the order of the nucleon Fermi momentum $p_F$. Provided $p_F < \Lambda_b$, $\chi$EFT
calculations of strongly interacting matter can, in principle, be improved to any desired accuracy. 
The errors at a given order are calculated by estimating the magnitude of the terms of the next higher order. 
In this way, a prior distribution with probabilities assigned to each of its members can be obtained. 

\begin{figure}[ht]		
\centerline{\includegraphics[width=0.60\columnwidth]{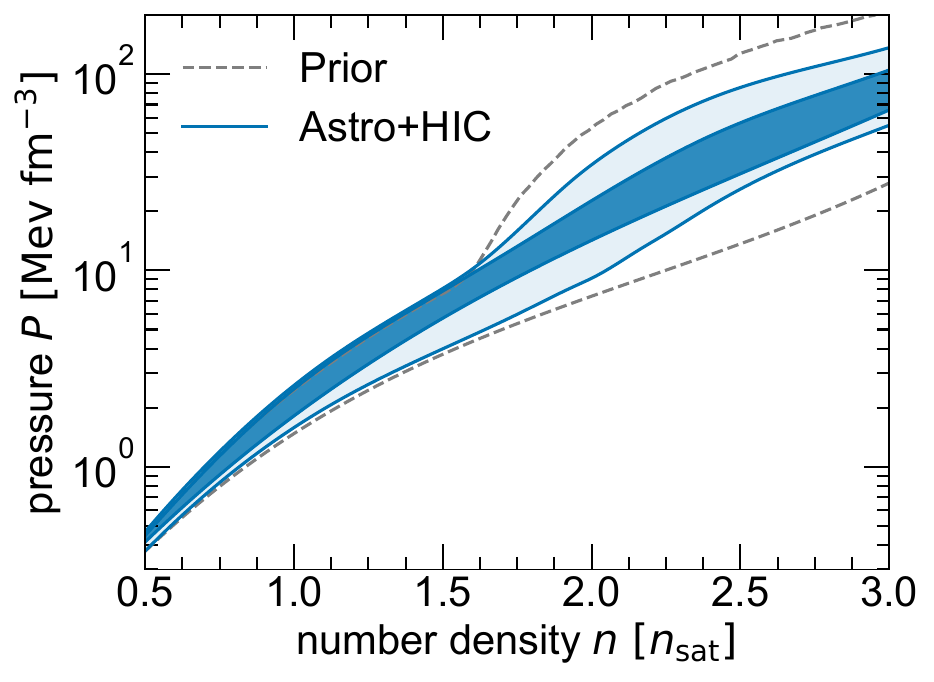}}
  \caption{Constraints on the equation of state of neutron-star matter represented as the evolution of the
pressure as a function of baryon number density when combining data from multi-messenger neutron-star observations
(Astro) and from heavy-ion collisions (HIC). The shading corresponds to the 95\% and 68\% credible intervals. 
The dashed lines represent the
95\% credible interval of the prior generated with $\chi$EFT and the requirement that a maximum neutron-star mass of at
least 1.9 solar masses is supported by all candidate equations of state
(adapted from Ref.\protect\cite{huth22}).
}
\label{fig:huth}
\end{figure}

In the work of Huth {\it et al.}\cite{huth22}, chiral effective field theory is used to cover the density interval between
the crust-core transition at about $0.5 n_{\rm sat}$ and $1.5 n_{\rm sat}$ which is argued to be safely below the breakdown limit.
The extrapolation of the prior distribution to higher densities 
is achieved with parameter-free extension schemes,\cite{hebeler13,tews18} constrained by requiring that
the speed of sound $c_s$ obeys causality ($c_s < c$ with $c$ denoting the speed of light) and guarantees stability 
($c_s \ge 0$ with $c_s^2 = \partial p/\partial \epsilon$, i.e. 
the increase of the pressure with energy density). With the additional requirement
that neutron stars with at least 1.9 solar masses are supported,
a radius $R_{1.4} = 11.96 \pm 1.18$~km (95\%) for a 1.4 solar-mass neutron star is predicted by the prior distribution. 

Using Bayesian inference, the results available for the masses and radii of neutron stars, results obtained from the 
observation of the neutron star mergers GW170817 and GW190425 and from the kilonova AT2017gfo are combined 
with the information obtained from heavy-ion collisions (Figs.~\ref{fig:wang} and \ref{fig:russotto_fig18}). 
Partly new software is used to deduce the information on the equation
of state contained in the gravitational wave signal of GW170817 and its electromagnetic counterpart.
The density distributions of the sensitivity of the elliptic flow results and the dependence of the ASY-EOS result on 
the assumed $E_{\rm sym}(\rho_0)$ were taken into account.

The resulting equation of state given in the form of pressure versus density is displayed in Fig.~\ref{fig:huth}. 
The distribution is significantly narrower than that obtained by only using astrophysical 
observations (Fig.~\ref{fig:legred}), in particular at densities below $2 n_{\rm sat}$. 
The final result for the radius of a 1.4 solar-mass neutron star is $R_{1.4} = 12.01 \pm 0.78$~km (95\% credibility). 
The importance of the choice of the prior becomes evident in the result $R_{1.4} = 12.56 \pm 1.07$~km obtained with a lower
breakdown density of $1.0 n_{\rm sat}$. The uncertainty is increased as expected but the value for the radius is larger by 
$\approx 500$~m if the $\chi$EFT prediction for the interval 1.0 to $1.5 n_{\rm sat}$ is ignored. 
It is also apparent in Fig.~\ref{fig:huth} that the final posterior distribution touches the upper limit of the prior 
distribution in this interval, indicating that information provided by astrophysical observations and heavy-ion collisions 
at supra-saturation densities deviates from the $\chi$EFT prediction as used in this work.
As noted earlier by Essick {\it et al.}, NICER observations suggest that the equation of state stiffens 
relative to $\chi$EFT predictions at or slightly above nuclear saturation density.\cite{essick20}

An alternative approach was presented by Tsang {\it et al.}\cite{tsang24}. The method of metamodeling is used for generating 
the prior distribution of equations of state. For constraining the parameters, 
the $\chi$EFT of Drischler {\it et al.}\cite{drischler21} and a selected list of laboratory data is employed, including
the results obtained with PREX-2\cite{adhikari21}, the S$\pi$RIT data presented by Estee {\it et al.},\cite{estee21}
and the ASY-EOS elliptic-flow result.\cite{russotto16}. Astrophysical results from NICER and LIGO/Virgo are then used to
inform the so obtained prior distribution. The radius $R_{1.4} = 12.9 \pm 0.5$~km (68\%) calculated from the
resulting posterior appears at the upper end of values obtained from recent combined analyses. Upper limits
reported earlier, after the discovery of $\approx 2$ solar-mass neutron stars, reach further up to nearly 14 km.\cite{hebeler13} 
There is a clear correlation between the apparently large radius $R_{1.4}$ and the laboratory constraints used
for generating the prior which all favor larger slope parameters $L$. 

Figure~\ref{fig:cozma_fopi} presents a comparison of the pressure versus density relations
obtained in three studies, Legred {\it et al.}\cite{legred21}, 
Huth {\it et al.}\cite{huth22}, and Tsang {\it et al.}\cite{tsang24}. 
Two of them were already shown and introduced in
Figs.~\ref{fig:legred} and~\ref{fig:huth}, and their agreement at high density and the effect of additional constraints 
at lower densities were addressed. We observe that the contours of the two combined studies\cite{huth22,tsang24}
(magenta and green) are practically mutually exclusive up to 1.5 times saturation density, 
even at the 95\% confidence limits displayed in the figure.
The different constructions of the priors and the choices of additional information relating to lower densities
are responsible for this significant effect. This observation was already reported in the White Paper of 
Sorensen{\it et al.}\cite{sorensen24} in which it was also pointed out that the $\chi$EFT result of 
Drischler {\it et al.}\cite{drischler21} 
stiffens with density and, at $1.5 n_{\rm sat}$, is considerably higher than the combined result of Huth {\it et al.}
who, for generating the prior distribution, used the softer $\chi$EFT results of Lynn {\it et al.}\cite{lynn16} 
up to density $1.5 n_{\rm sat}$. 
It is evident that $\chi$EFT predictions at these densities
are not yet sufficiently robust (see, e.g., Refs.\cite{drischler21a,huth21,tews24})  
and that the mode of using them has a strong influence on the finally obtained posterior distributions of equations of state.

The work of Huth {\it et al.} has been extended by Koehn {\it et al.} who have added more data sets of 
astrophysical observations and laboratory measurements.\cite{koehn25} 
Altogether, 20 different types of input are used for informing a prior constructed with the technique of metamodeling. 
$\chi$EFT is used as information for densities extending up to between 1 and $2 n_{\rm sat}$.
The outcome $R_{1.4} = 12.20 \pm 0.50$~km (95\%) is slightly larger than that of 
Huth {\it et al.}\cite{huth22} and may be considered as resulting from the presently most complete analysis. 
Each of the 20 sources of input information is critically discussed in the report.

The analysis of Pang {\it et al.} does not make use of information from heavy-ion collisions.\cite{pang24a} 
With a prior obtained with $\chi$EFT and exclusively astrophysical inputs, 
the value $R_{1.4} = 11.98 \pm 0.40$~km at 90\% credible interval is obtained. It carries a comparatively small error
and, at present, constitutes the lower limit of the interval spanned by combined analyses. 
It is evident that ignoring information coming from heavy-ion collisions
and relying on $\chi$EFT up to $2 n_{\rm sat}$ leads to softer results for the neutron-star-matter equation of state.

\section{Discussion and Outlook}
\label{sec:discussion}

When this article was near completion, several interesting new reports have appeared. The analysis groups within
the NICER Collaboration published improved results for the stars under observation
which had already been announced in their earlier publications.\cite{riley21,miller21}. Salmi {\it et al.}\cite{salmi24} and
Dittmann {\it et al.}\cite{dittmann24} were able to include additional observational data in their analyses and profited from
improvements in the light-curve analysis of the heaviest known neutron star J0740+6620. Their results for the equatorial radius
of this star, $R = 12.49 ^{+ 1.28} _{- 0.88}$~km and $R = 12.76 ^{+ 1.49} _{- 1.02}$~km (both 68\%), respectively, are fully consistent 
with each other and with the expectation of smaller errors following from additional observational input. They will serve as 
strong anchor points for the nuclear equation of state at high density, in particular also excluding significant hyperon
components in neutron stars.\cite{lonardoni15,logoteta19,ye24,weise24}

\begin{figure}[ht]		
\begin{minipage}{6.2cm}
\centerline{\includegraphics[width=6.5cm]{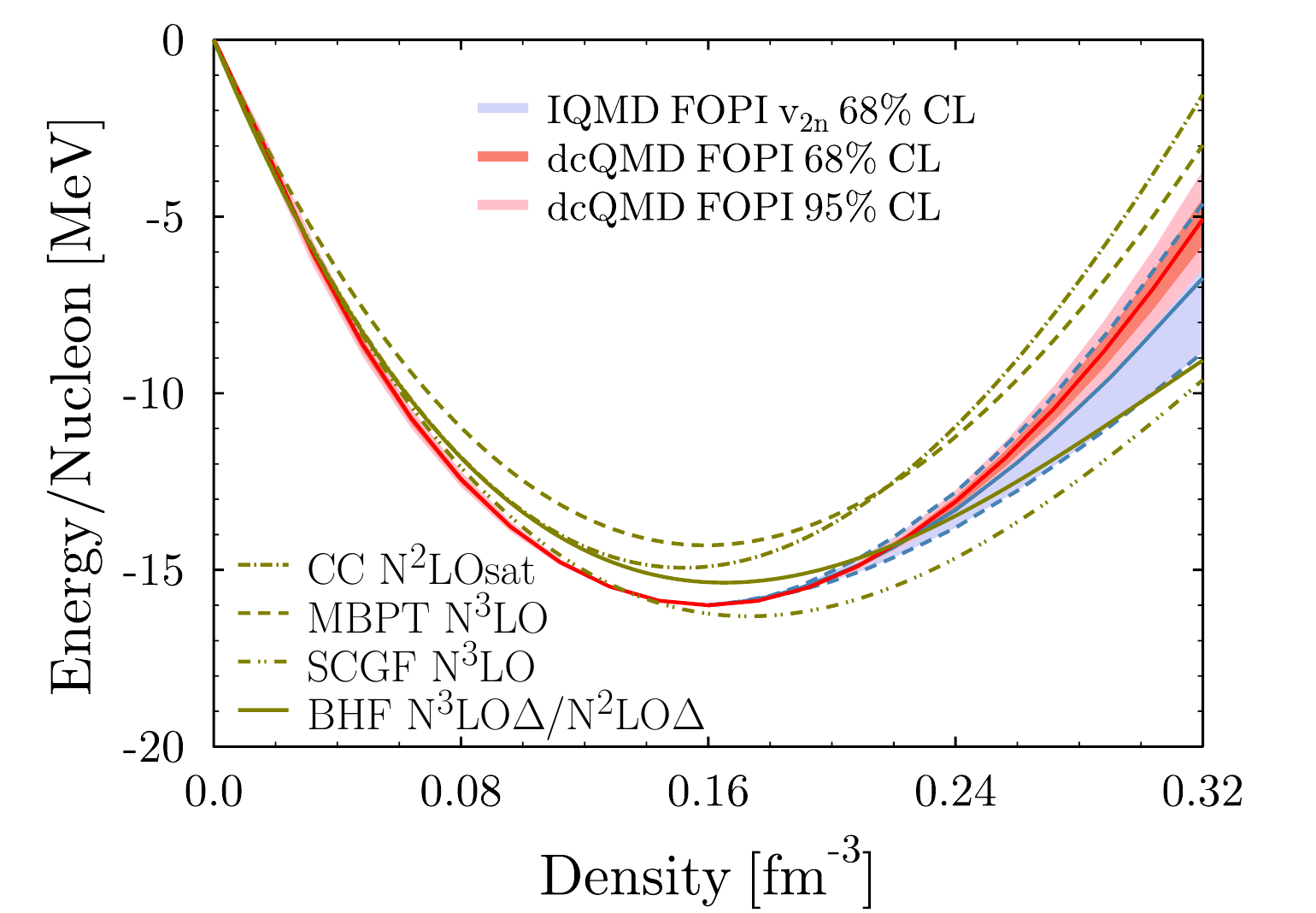}}
\end{minipage}
\begin{minipage}{6.2cm}
\centerline{\includegraphics[width=6.5cm]{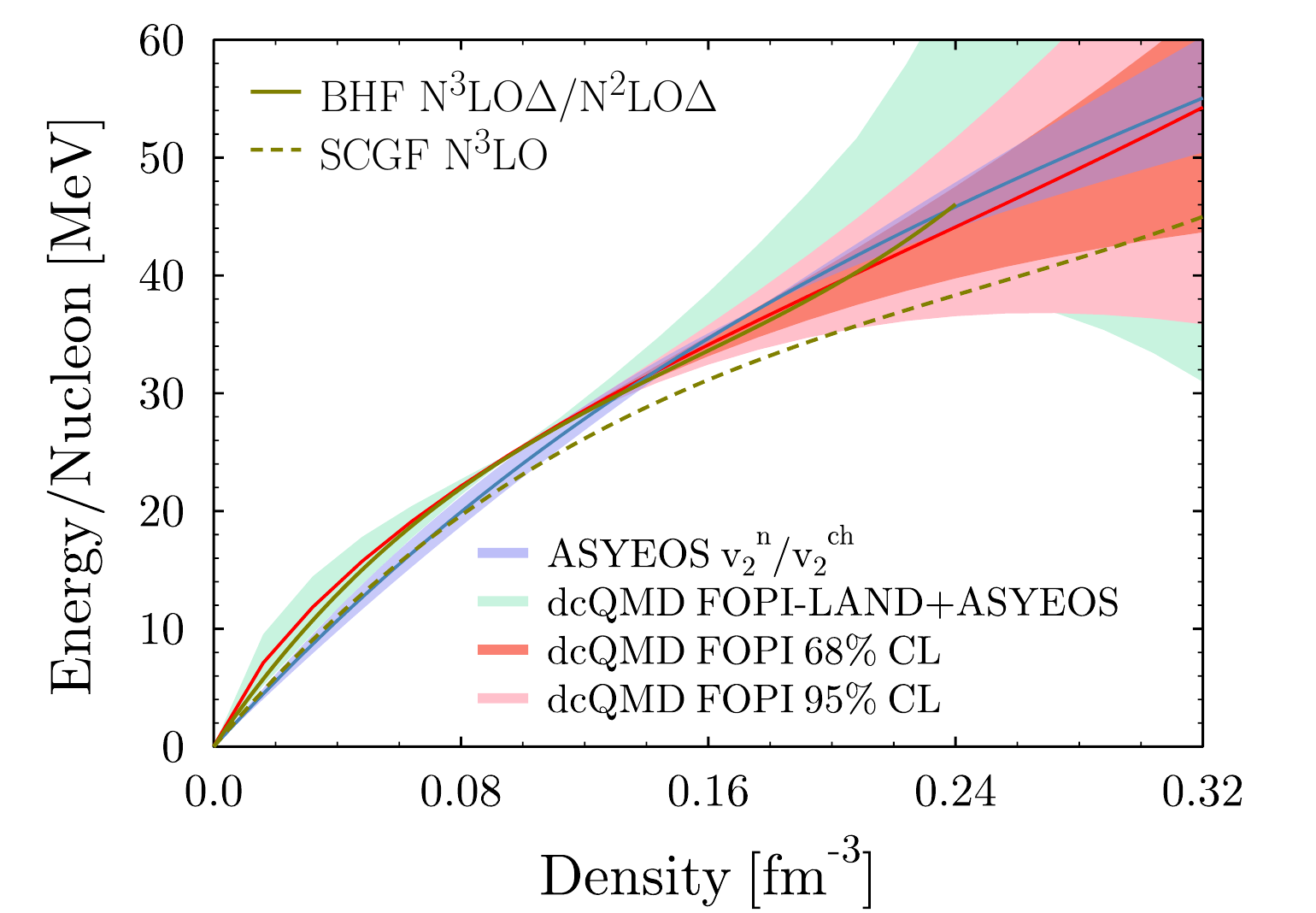}}
\end{minipage}
  \caption{Constraints obtained in Ref.\protect\cite{cozma24} (dcQMD FOPI, red lines and contours)
   for the density dependence of the equation of state of symmetric nuclear matter (left panel) 
   and for the symmetry energy (right panel) in comparison with other analyses of data from
   heavy-ion collisions and {\it ab-initio} calculations (black lines); see text for more explanations and 
   Ref.\protect\cite{cozma24} for references to the selected set of theoretical predictions
   (reprinted with permission from Ref.\protect\cite{cozma24}, copyright \textcopyright~2024 by the American Physical Society).
}
\label{fig:cozma12}
\end{figure}

Most recently, a very interesting observation was published by the NICER Collaboration,\cite{choudhury24}
reporting a radius $R = 11.36 ^{+ 0.95} _{- 0.63}$~km (68\%) for the millisecond pulsar J0437-4715 whose mass 
$M = 1.418 \pm 0.044$ is known with high precision.\cite{reardon24} PSR J0437-4715 is the brightest
pulsar on the NICER list of candidates for pulse profile modeling and with a distance of about 500 light years 
much closer than PSR J0030+0451 (1060 light years), the first star whose radius was reported by the NICER 
Collaboration.\cite{riley19,miller19,vinciguerra24} With its mass of 1.4 solar masses, PSR J0437-4715 qualifies as a canonical
neutron star and its radius may thus serve as a direct measurement to be compared to existing analysis results for this quantity. 
The central value of 11.4 km is lower than the interval 12 to 13 km appearing as favored by the combined analyses discussed
in the previous section. With its error of nearly 1~km (68\%) it extends, however, deeply into this region. 
In a first combined analysis\cite{rutherford24} including the new results,\cite{salmi24,dittmann24,choudhury24} radii
$R_{1.4}$ between 12.01 and 12.28 km, depending on the chosen extension model, were obtained with errors of $\approx 0.7$~km (95\%).
In Refs.\cite{malik24,brandes25} it is found that the new NICER measurement, if included, leads to an average radius
reduction of $\approx 0.1$~km to 0.2 km in the posterior of the obtained mass-radius relationships.
The reported radius of J0437-4715 is thus not in contradiction with previous combined analyses and their results.

The densities in the interior of the canonical 1.4-solar-mass neutron stars extend to $\approx 3-4 n_{\rm sat}$\cite{abbott18} and 
their radii are thus dominated by the pressure at lower densities near and around $2 n_{\rm sat}$. This is the upper 
end of the density range accessible with heavy-ion collisions which emphasizes their role in clarifying the 
pressure versus density relation in this important interval of $1-2 n_{\rm sat}$ (Fig.~\ref{fig:cozma_fopi}). 
Transport models are indispensable for the theoretical description of heavy-ion collisions and for extracting information
from their high-density stages; continuously improving them and controlling their consistency is thus an important task.

As an illustration of the current situation, we present in Fig.~\ref{fig:cozma12} the results reported in 
Ref.\cite{cozma24} in comparison with selected previous work. 
For symmetric nuclear matter (left panel), the energy per nucleon deduced from
the FOPI stopping and flow data and characterized by $K_0 = 230 \pm 11$~MeV (in red) touches the result of 
Le F\`{e}vre {\it et al.}\cite{lefevre16} (IQMD FOPI, $K_0 = 190 \pm 30$~MeV) from above and is in good agreement 
with the values $K_0 = 220 \pm 40$~MeV of Wang {\it et al.} (Fig.~\ref{fig:wang})
and $K_0 = 200 \pm 25$~MeV as adopted in the combined analysis of Huth {\it et al.}\cite{huth22} (all errors represent 68\%
confidence limits). The symmetric-matter incompressibility of this magnitude seems fairly robust 
and is found to be favored also by recent similar studies\cite{taras24,kireyeu24} based on a wider selection of data
from heavy-ion collisions, including the high-statistics data of the HADES Collaboration.\cite{kardan16,hades23} 

The obtained density dependence of the symmetry energy (red line and contours in the right panel) is of particular 
interest because it reflects the
information contained in the isotopic dependences of light charged-particle cross sections and flows as discussed in
Section~\ref{sec:interp}. With increasing density, it follows closely the original ASY-EOS result\cite{russotto16} as well as 
the central value obtained in the subsequent analysis\cite{cozma18} of the FOPI-LAND and ASY-EOS data. In the latter case,
the uncertainty grows rapidly toward $2 n_{\rm sat}$ because the contribution of the less well determined $K_{\rm sym}$ 
parameter becomes important (Eq.~(\ref{eq:taylor})). This work had demonstrated that it is possible to determine $L$ and
$K_{\rm sym}$ independently but that additional experimental data will be needed for achieving more satisfactory
results. 

New results may be expected from presently performed experiments 
at the RIKEN and GSI/FAIR laboratories. Data taking by the S$\pi$RIT Collaboration at RIKEN has been completed in 2024,
with emphasis given to collecting the higher statistics found necessary for analyzing pion ratios at high transverse meomentum
(Fig.~\ref{fig:estee} in Section~\ref{sec:HIC}). The density regime covered with this method is estimated to be 
around $1.5 n_{\rm sat}$.\cite{tsang24} The ASY-EOS Collaboration is scheduled for data taking at the GSI laboratory in 2025 with
the aim to extend the probed density interval up to about $2 n_{\rm sat}$ by measuring an excitation function of elliptic flows for 
incident energies up to 1 GeV/nucleon.\cite{ASYEOSII} The expected mass resolution for light charged particles and improved
count rate statistics will enhance the significance of the measured flow ratios (Fig.~\ref{fig:cozma3} in Section~\ref{sec:interp}).

\section{Conclusion}
\label{sec:conclusion}

Recent analyses based on combining information obtained from nuclear theory, heavy-ion 
collisions and astrophysical observations have confined the obtained radii of the canonical 1.4-solar-mass 
neutron star to values between 12 and 13 km. Typical errors are still of the order of 0.9 km (95\%) but can be as low as
0.5 km (95\%) or 0.4 km (90\%) as reported in the most recent analyses of Koehn {\it et al.}\cite{koehn25} and Pang
{\it et al.}\cite{pang24a}, respectively.  

At the time of its observation, it was not expected that GW170817 will remain the only neutron-star merger of its kind 
for such a long time.\cite{chaves19} Observing run 04b of LIGO/Virgo/KAGRA which started in April 2024  
is scheduled to continue until June 2025.\cite{abac24} 
The LIGO and Virgo observatories are presently online with their expanded reach due to higher sensitivity. 
The Kamioka Gravitational Wave Detector KAGRA is expected to join as well. The consistent picture that has emerged from the GW170817 multi-messenger
observations and their interpretations makes it unlikely, however, that any new observation of comparable significance will be 
severely contradicting existing results. Their confirmation and possible improvement is still very important, 
and a reduction of remaining uncertainties can be expected.

Altogether, one may expect that the precision of the result of Legred {\it et al.}\cite{legred21} (Fig.~\ref{fig:legred})
will be improved without changing the overall behavior. In particular, the data now available for PSR J0740+6620 represent
a very strong reference point for neutron-star matter up to high densities. Many other sources of astrophysical information 
as used by Koehn {\it et al.}\cite{koehn25} may profit from continuing analyses and observations. 
The statistical weight of information resulting from astrophysical observations is thus expected to increase.

At low densities, the precise knowledge of the symmetry energy at 2/3 saturation density may be considered as the second robust
reference point. Open issues at lower densities are the precision of the symmetry energy at saturation density and the pressures 
in the interval up to twice that value which, however, are accessible with laboratory experiments. We may, therefore, 
expect that heavy-ion experiments presently prepared and conducted at the RIKEN and GSI/FAIR laboratories have the potential 
of reducing the still existing uncertainties in this density interval. 
Continuing activities in the field of transport theory\cite{wolter22,cozma24} will be essential for achieving this goal.\\

{\bf Acknowledgment}\\
The authors would like to thank C. A. Raithel, R. Essick, N. Chamel, Yongjia Wang, W. G. Lynch, M. B. Tsang and 
P. T. H. Pang for providing graphic material and for stimulating communications. The continuing fruitful scientific exchange
within the ASY-EOS Collaboration\cite{ASYEOSII} is gratefully acknowledged.


\end{document}